\documentclass[
 reprint,
secnumarabic,
 amssymb, amsmath,
 aip,apr,
groupedaddress,
frontmatterverbose,
]{revtex4-1}
\usepackage{graphicx}
\usepackage{docs}%
\usepackage{bm}%
\usepackage[colorlinks=true,linkcolor=blue]{hyperref}%
\expandafter\ifx\csname package@font\endcsname\relax\else
 \expandafter\expandafter
 \expandafter\usepackage
 \expandafter\expandafter
 \expandafter{\csname package@font\endcsname}%
\fi
\usepackage{url}
\begin{document}
\title{Encoding Neural and Synaptic Functionalities in Electron Spin: A Pathway to Efficient Neuromorphic Computing}

\author{Abhronil Sengupta}
\email{asengup@purdue.edu}
\author{Kaushik Roy}

\affiliation{School of Electrical \& Computer Engineering, Purdue University, West Lafayette, Indiana 47907, USA}

\begin{abstract}
\small{Present day computers expend orders of magnitude more computational resources to perform various cognitive and perception related tasks that humans routinely perform everyday. This has recently resulted in a seismic shift in the field of computation where research efforts are being directed to develop a neurocomputer that attempts to mimic the human brain by nanoelectronic components and thereby harness its efficiency in recognition problems. Bridging the gap between neuroscience and nanoelectronics, this paper attempts to provide a review of the recent developments in the field of spintronic device based neuromorphic computing. Description of various spin-transfer torque mechanisms that can be potentially utilized for realizing device structures mimicking neural and synaptic functionalities is provided. A cross-layer perspective extending from the device to the circuit and system level is presented to envision the design of an All-Spin neuromorphic processor enabled with on-chip learning functionalities. Device-circuit-algorithm co-simulation framework calibrated to experimental results suggest that such All-Spin neuromorphic systems can potentially achieve almost two orders of magnitude energy improvement in comparison to state-of-the-art CMOS implementations.}
\end{abstract}

\maketitle
\tableofcontents

\makeatletter
\let\toc@pre\relax
\let\toc@post\relax
\makeatother 

\section{Introduction}
Although the brain is not yet fully understood, neuromorphic computing that attempts to emulate some facets of its functionalities and inter-connectivity, are becoming increasingly popular on machine learning tasks, and are surpassing humans at multiple cognitive tasks more than ever before. For instance, recently Google DeepMind beat a professional human champion at a $19 \times 19$ Go board game \cite{silver2016mastering}. The key inspiration behind the development of algorithms and computing paradigms with high degree of bio-fidelity is driven by the expectation that by emulating some attributes of the human brain, we would be able to approach the brain's highly efficient and low-power cognitive abilities. For instance, implementation of bio-realistic ``spiking" neural computing paradigms have recently enabled low-power event-driven neuromorphic hardware equipped with on-chip local spike-timing dependent synaptic learning functionalities.

While these neuro-inspired computing models are still implemented in von-Neumann architectures consisting of Boolean logic and memory circuits, the brain's ``computing fabric" is highly parallel, interconnected and enabled with in-situ synaptic memory storage. Further CMOS transistors, that form the underpinnings of current computing systems, are on-off switches that are naturally suited for Boolean computing but may not inherently map to the ``computational primitives" of neuro-mimetic algorithms. Limited by this mismatch between the computational units and the underlying hardware, CMOS based neuromorphic architectures consume resources and power that are orders of magnitude higher than that involved in the biological brain \cite{adee2009ibm}. Bridging this gap necessitates the exploration of devices, circuits and architectures that provide a better match to biological processing and which require a significant rethinking of traditional von-Neumann based computing.

\begin{figure}[ht!]
\centering
\includegraphics[width = 3.4in ]{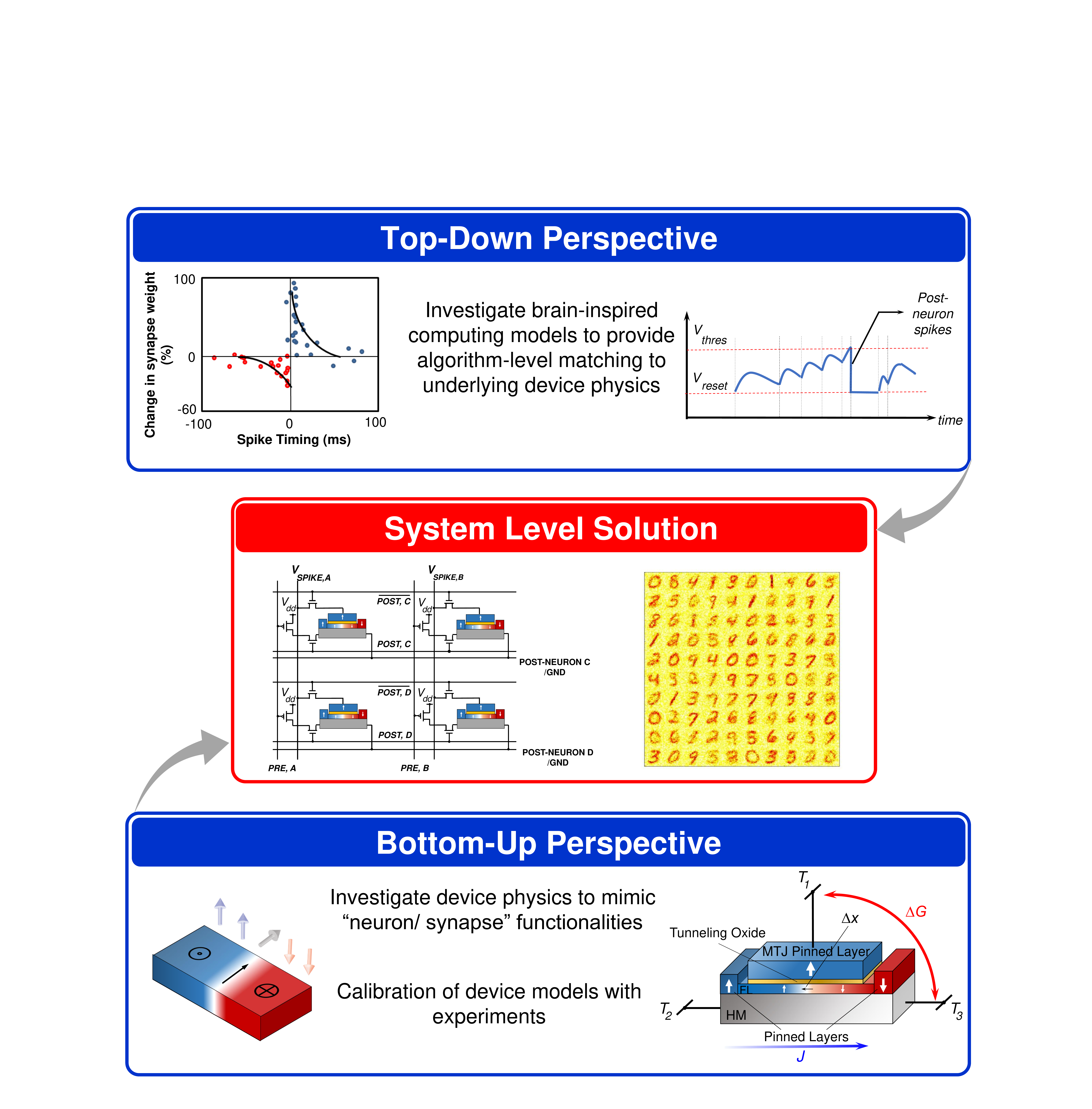}
\caption{Cross-layer research effort across the stack of materials, devices, circuits and algorithms to provide system-level solutions for enabling cognitive intelligence. A ``top-down" perspective to provide algorithm-level matching to the underlying device physics of spintronic devices is complemented by a ``bottom-up" approach where recent experiments in spintronics are leveraged to propose device structures that can directly mimic neural and synaptic functionalities.}
\label{overview}
\end{figure}

While usage of spintronic devices in memory applications have achieved maturity and is close to the market \cite{fongspin}, recent experiments in domain wall motion based devices \cite{sengupta2016vision,sengupta2015hybrid} and probabilistic switching characteristics of scaled nanomagnets \cite{sengupta2015magnetic,srinivasan2016magnetic} are revealing immense possibilities of implementing a plethora of neural and synaptic functionalities by single spintronic device structures that can be operated at very low terminal voltages. Simple engineering of the device dimensions or biasing region of the operating transistors can enable the emulation of functionalities that can range from neuron spiking behavior to synaptic learning abilities in the same magnetic stack. While other emerging devices such as resistive memories have also been explored for neuromorphic computing, they are limited by the variety of neural or synaptic functionalities that they can emulate along with high energy requirements for programming \cite{jackson2013nanoscale,kuzum2011nanoelectronic} (which is an essential component of learning and neural inference). The prospect of large improvements in integration density and energy consumption and concurrently providing in-memory computing possibilities (due to their inherent non-volatility) can potentially make spintronic devices a promising path towards realizing ``brain-like" nanoelectronic computing. This article attempts to provide a multi-disciplinary perspective across the entire stack of materials, devices, circuits, systems and algorithms where understanding of the underlying device physics of spintronic devices (``bottom-up approach") is complemented by efforts to adapt neuromorphic computing models to the unique characteristics of spintronic devices (``top-down approach") to construct cognitive networks of interconnected spintronic neural and synaptic components (Fig. \ref{overview}). 

\section{Spintronic Devices: Underlying Physical Phenomena}
Several spintronic device structures have been proposed in literature to mimic different neuronal or synaptic functionalities. However, in order to understand the mapping of biological functions to the operation of such spin devices, an understanding of the underlying physical phenomena is necessary. This section provides a brief overview of major spin-torque effects in nanomagnets that can be engineered to realize such neuromimetic computations.

The two main physical phenomena that are exploited to construct neuromimetic spin devices are the spin-torque effect (``write" mechanism) and the Tunneling Magneto-Resistance or the TMR effect (``read" mechanism). The manipulation of magnetization state without the assistance of any external magnetic field through spin-transfer torque effect was first predicted by Slonczewski \cite{slonczewski1996current} and Berger \cite{berger1996emission} in 1996. Several experiments demonstrating spin-transfer torque induced magnetization reversal have been demonstrated henceforth \cite{myers1999current,grollier2001spin,yuasa2004giant}. On the other hand, sensing the magnetization state through the TMR effect was first experimentally observed by Julli\'{e}re in 1975 in Fe/Ge-O/Co stacks \cite{julliere1975tunneling}.

\subsection{Device Fundamentals}
A nanomagnet is characterized by two collinear but oppositely directed stable magnetization directions, termed as the ``easy'' axis, such that in the absence of any external perturbation (magnetic field or input spin current) the magnetization would relax to either of the stable magnetization states. The stability of the magnet in the presence of thermal noise is maintained by virtue of a barrier height, $E_{B}$, that is determined by the uniaxial anisotropy, $K_{u2}$, of the magnet as \cite{sun2005tuning},
\begin{equation}
E_{B} = K_{u2}V
\end{equation}
where $V$ is the volume of the magnet. The lifetime of the magnet in absence of thermal agitation is related exponentially to the magnitude of the barrier height. For instance, a barrier height of $40K_{B}T$ ($K_{B}$ is Boltzmann constant) ensures a magnet lifetime of $\sim 7.4$ years \cite{sun2005tuning}.

The uniaxial anisotropy of the magnet, and hence the direction of magnet ``easy-axis", can be in-plane (IMA) when shape anisotropy dominates the resultant anisotropy of the magnet \cite{fongspin,driskill2011latest,jeong20030}. In this case, the magnet cross-sectional area would be an ellipse with the ``easy-axis'' being in the direction of the longer dimension. In contrast, in perpendicular magnetic anisotropy (PMA) materials, the magnetocrystalline anisotropy dominates over the shape anisotropy in order to make the out-of-plane direction as the ``easy-axis'' direction \cite{fongspin,ikeda2010perpendicular,gajek2012spin}. Hence, PMA magnets are usually of circular cross-sectional area.
\begin{figure*}
\centering
\includegraphics[width = 5.8in ]{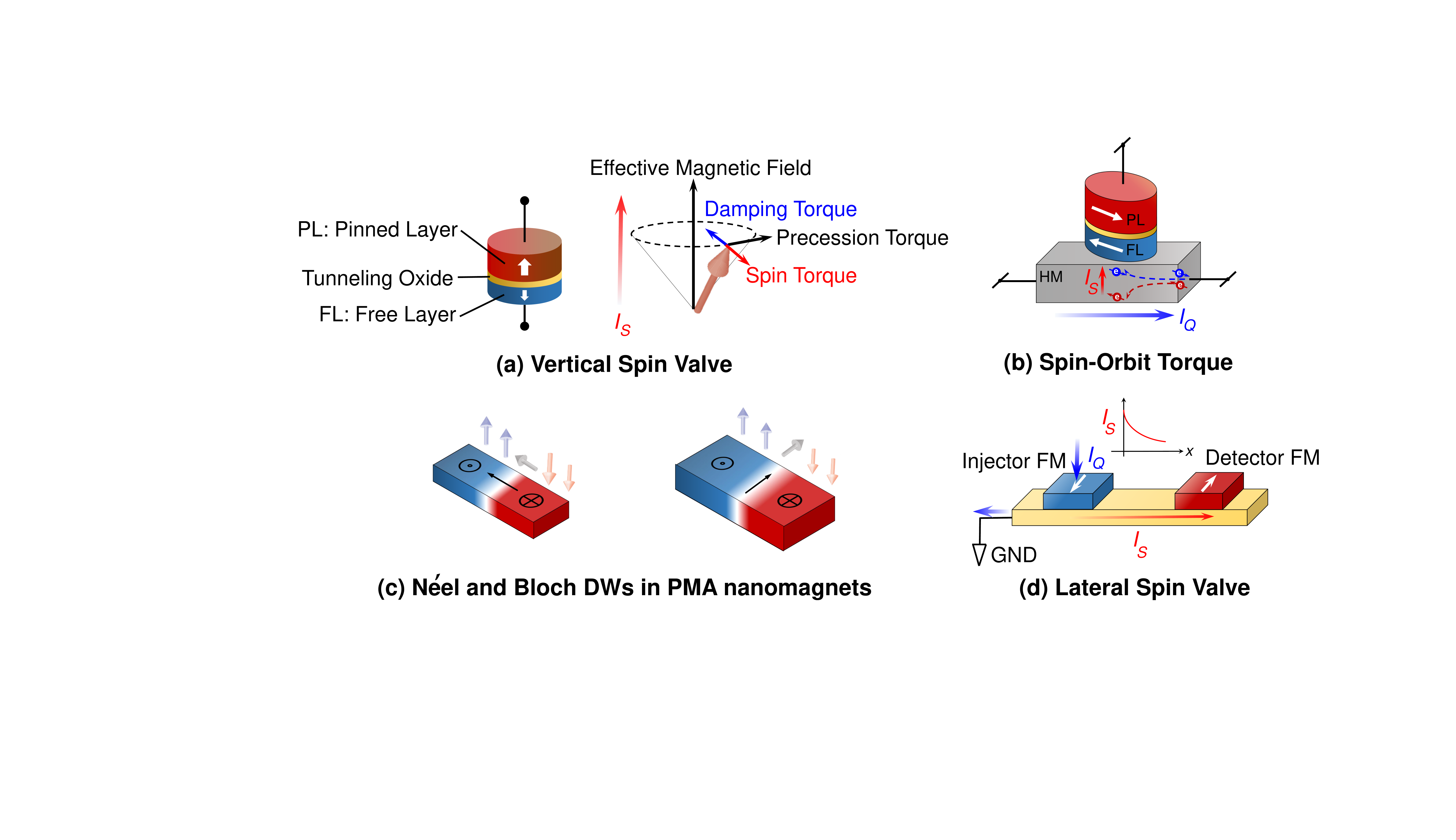}
\caption{(a) Vertical Spin Valve: A Magnetic Tunnel Junction consists of two ferromagnets, namely the ``free" layer (FL) and the ``pinned" layer (PL) separated by a tunneling oxide barrier. The magnetization dynamics evolves under the influence of the damping torque, precession torque and spin-torque due to an input spin current, $I_S$. (b) N\'{e}el and Bloch domain walls (DWs) observed in narrow and wider nanostrips with Perpendicular Magnetic Anisotropy (PMA) respectively. (c) Spin-orbit torque is generated on a nanomagnet due to charge current flow ($I_Q$) through an underlying Heavy Metal (HM) layer due to spin-Hall effect. (d) Lateral Spin Valve based structure where an injector and detector ferromagnet are located on top of a non-magnetic channel. The detector ferromagnet can be switched due to non-local spin-torque effect exerted by charge current flowing through the injector magnet to the ground contact lying beneath the magnet. The magnitude of the injected spin current, $I_S$, reduces exponentially with the distance between the injector and detector FMs.}
\label{prelim}
\end{figure*}

In order to read the magnetization state of the nanomagnet, a Vertical Spin Valve (VSV) structure is utilized as shown in Fig. \ref{prelim}(a). It is referred to as the Magnetic Tunnel Junction (MTJ) \cite{yuasa2004giant,julliere1975tunneling,parkin2004giant} where a thin oxide acts as the tunneling barrier between two nanomagnets. The resistance of the MTJ depends on the relative orientation of the magnetization directions of the two nanomagnets. In order to provide a reference, the magnetization of one of the magnets is pinned to a particular direction (usually achieved by coupling to an antiferromagnetic layer), $\widehat{\textbf{m}}_P$, while the magnetization of the other layer, $\widehat{\textbf{m}}$, can be determined by the resistance of the MTJ stack. The two layers are referred to as the ``pinned'' layer (PL) and ``free'' layer (FL) respectively. The difference in resistance of the MTJ with relative magnetic orientations of the FL and PL can be explained from the concept of ``spin-filtering'' \cite{fongspin,inoue2009gmr}. When $\widehat{\textbf{m}}_P$ and $\widehat{\textbf{m}}$ are parallel to each other (Parallel configuration: P), electrons with that corresponding spin orientation can easily tunnel through the oxide since the filled states in the band structure of one contact corresponding to that particular spin orientation is well matched to empty states for the same spin in the other contact. On the contrary, when $\widehat{\textbf{m}}_P$ and $\widehat{\textbf{m}}$ are oppositely directed (Anti-Parallel configuration: AP), the band structures of either spin configuration are not well-matched for the two contacts, thereby resulting in higher resistance. The metric utilized to measure the difference between the P ($R_P$) and AP ($R_{AP}$) MTJ resistances is referred to as the Tunneling Magnetoresistance Ratio (TMR) defined as,
\begin{equation}
TMR = \frac{R_{AP}-R_{P}}{R_{P}}\times 100\%
\end{equation}
It is worth noting here that the MTJ P and AP resistances are a function of the oxide thickness and applied voltage across the MTJ which can be formulated using the Non-Equilibrium Green's Function based transport simulation framework \cite{fong2011knack}. 

The discussion so far has been limited to sensing the magnetization state of a nanomagnet. Let us now discuss the mechanism of manipulating the magnetization direction of a magnet. One of the most common mechanisms is by passing a charge current through the MTJ stack due to spin-transfer torque effect \cite{slonczewski1996current,berger1996emission,myers1999current,grollier2001spin,yuasa2004giant}. When charge current flows from the FL to the PL, electrons are injected into the FL from the PL that are spin-polarized in the direction of $\widehat{\textbf{m}}_P$. The magnitude of injected spin current is determined by the polarization of the magnet. Hence the injected spins attempt to orient the FL in the direction of $\widehat{\textbf{m}}_P$. For a sufficient magnitude of the current flowing from the FL to the PL, the MTJ is switched to the P configuration. On the other hand, when current flows from the PL to the FL, the FL attempts to inject spins into the PL. However, due to ``spin-filtering'', only electrons with spin parallel to $\widehat{\textbf{m}}_P$ can tunnel easily to the PL from the FL. Hence the remaining spins anti-parallel to $\widehat{\textbf{m}}_P$ remain in the FL and exert a torque to orient the MTJ in the AP state. 

The temporal evolution of magnetization dynamics can be described by Landau-Lifshitz-Gilbert equation \cite{sun2000spin} with additional terms to account for the effect of spin-transfer torque \cite{slonczewski1989conductance} as follows,
\begin{equation}
\label{llg}
\frac {d\widehat {\textbf {m}}} {dt} = -\gamma(\widehat {\textbf {m}} \times \textbf {H}_{eff})+ \alpha (\widehat {\textbf {m}} \times \frac {d\widehat {\textbf {m}}} {dt})+\frac{1}{qN_{s}} (\widehat {\textbf {m}} \times \textbf {I}_s \times \widehat {\textbf {m}})
\end{equation}
where $\widehat {\textbf {m}}$ is the unit vector of FL magnetization, $\gamma= \frac {2 \mu _B \mu_0} {\hbar}$ is the gyromagnetic ratio for electron, $\alpha$ is Gilbert\textquoteright s damping ratio, $\textbf{H}_{eff}$ is the effective magnetic field, $N_s=\frac{M_{s}V}{\mu_B}$ is the number of spins in free layer of volume $V$ ($M_{s}$ is saturation magnetization and $\mu_{B}$ is Bohr magneton), and $\textbf{I}_{s}$ is the input spin current generated by the HM underlayer. Thermal noise is included by an additional thermal field \cite{scholz2001micromagnetic}, $\textbf{H}_{thermal}=\sqrt{\frac{\alpha}{1+\alpha^{2}}\frac{2K_{B}T_{K}}{\gamma\mu_{0}M_{s}V\delta_{t}}}G_{0,1}$, where $G_{0,1}$ is a Gaussian distribution with zero mean and unit standard deviation, $K_{B}$ is Boltzmann constant, $T_{K}$ is the temperature and $\delta_{t}$ is the simulation time-step.

In the absence of any input current stimulus, the magnet is subjected to a field-torque (that causes it to precess in the direction of the effective magnetic field) and a damping torque (that attempts to stabilize the magnet along the initial equilibrium state). The effective magnetic field includes any external applied field, magnetic uniaxial anisotropy field along with a thermal fluctuation field \cite{scholz2001micromagnetic,brown1963thermal} that lends a stochastic behavior to the switching process. The impact of input current on the magnetization dynamics is usually described by a Slonczewski-like torque \cite{slonczewski1989conductance} that acts in the plane of the damping torque and stabilizes the magnet along either of the two stable magnetization directions depending on the direction of the input spin current. Although some experiments have reported contributions from a field-like torque to the resultant spin-torque due to the input current \cite{matsumoto2011spin}, its magnitude is usually much less in comparison to the Slonczewski-like torque in tunneling junctions.

\subsection{Domain Wall Motion}
Mono-domain magnets where the entire FL magnetization is uniformly polarized can represent only two binary states. More than two states can be represented by multi-domain magnets that are fabricated with elongated shape to stabilize a transition region (termed as domain wall, DW) between two regions of opposite magnetic polarizations. The device state can be then represented by the position of the DW or the relative proportion of the two oppositely polarized magnetic domains. The manner of magnetization transition at the DW location depends on the anisotropy and shape of the magnet. While IMA nanowires are characterized by transverse (thin and narrow nanostrips) or vortex DWs (wider and thicker nanostrips) \cite{mcmichael1997head}, PMA materials exhibit N\'{e}el (narrow nanostrips) or Bloch DWs (wider nanostrips) \cite{torok1965transition}. Fig. \ref{prelim}(b) depicts the magnetic orientations of N\'{e}el and Bloch DWs observed in PMA magnetic strips. Current induced DW motion in the direction of electron flow was predicted \cite{berger1978low} and also observed in multiple experiments \cite{yamaguchi2005effect,beach2008current}. DW motion due to charge current flow through the magnet can be attributed to spin-torque generated due to local magnetization tracking of electrons flowing through the magnet.

\subsection{Spin-Orbit Torque}
Spin current generated by STT effect is always limited by the polarization strength of the injector magnet. Recent experiments on Insulator-Ferromagnet-Heavy Metal (I-FM-HM) multilayer structures have opened up the possibility of much greater spin injection efficiencies due to strong spin-orbit interaction (SOI) \cite{brataas2014spin} observed in such multilayer structures. When a charge current flows through the underlying HM, spin-orbit torque (SOT) is generated at the FM-HM interface. Although the cause of SOT can be attributed to two possible origins, namely the Rashba field due to structural inversion asymmetry \cite{miron2011fast} and the spin-Hall effect (SHE) \cite{hirsch1999spin}, we will consider SHE to be the dominant underlying physical phenomena for this text. As shown in Fig. \ref{prelim}(c), due to the flow of charge current through the HM, electrons with opposite spins scatter on the top and bottom surfaces of the HM. The spin-polarization is orthogonal to both the directions of charge current and injected spin current. These electrons experience spin-scattering repeatedly while travelling through the HM and thereby transfer multiple units of spin angular momentum to the FM lying on top. The magnitude of injected spin current density ($J_s$) is proportional to the magnitude of input charge current density ($J_q$), with the proportionality factor being defined as the spin-Hall angle \cite{hirsch1999spin} ($\theta_{SH}<1$). Hence, the input charge to spin current conversion is governed by the following relation,
\begin{equation}
I_s = \theta_{SH}.\left(\frac{W_{FM}}{t_{HM}}\right)I_q
\end{equation}
where $I_{s}$ and $I_{q}$ are the input spin current and charge current magnitudes respectively, $W_{FM}$ is the width of the FM lying on top of the HM, and $t_{HM}$ is the HM thickness. By ensuring $W_{FM}>>t_{HM}$, high spin injection efficiencies greater than $100\%$ ($I_s>I_q$) can be achieved. Typical HMs with high spin-orbit coupling under exploration are Pt, $\beta$-W and $\beta$-Ta. An important point to note is that the injected spins at the FM-HM interface have in-plane spin polarization due to SHE. Hence, SOT induced magnetization reversal is only possible for IMA magnets while an external magnetic field is required to switch PMA magnets in presence of SOT \cite{liu2012spin,yu2014switching,liu2012current}. Energy efficient SOT induced DW motion has been also observed in FM-HM bilayers in the presence of Dzyaloshinskii-Moriya interaction (DMI) \cite{dzyaloshinsky1958thermodynamic,moriya1960anisotropic}. The DMI effect can be modelled by including an additional field ($\textbf{H}_{DMI}$) in the calculation of the effective field $\textbf{H}_{eff}$ and is given by,
\begin{equation}
\textbf{H}_{DMI} = -\frac{2D}{\mu_{0}M_{s}}\left[\frac{\partial m_{z}}{\partial x}\widehat{x} + \frac{\partial m_{z}}{\partial y}\widehat{y} - \left(\frac{\partial m_{x}}{\partial x}+\frac{\partial m_{y}}{\partial y}\right )\widehat{z}\right ]
\end{equation}
where $D$ represents the effective DMI constant and determines the strength of DMI field in such multilayer structures. A positive sign of $D$ implies right-handed chirality and vice versa. The DMI effect is observed in PMA multilayers in the presence of spin-orbit coupling and broken inversion symmetry and results in the stabilization of N\'{e}el domain walls with a fixed chirality \cite{ryu2013chiral,ryu2014chiral,emori2014spin,emori2013current}. Such homochiral N\'{e}el DWs can be displaced by injected spins at the FM-HM interface due to charge current flowing through the underlying HM. Bloch DW motion in PMA multilayers with negligible DMI has been also achieved in the presence of an external magnetic field \cite{bhowmik2015deterministic}.

\subsection{Lateral Spin Valves}
Spin current injection can also occur in Lateral Spin Valve (LSV) structures, as depicted in Fig. \ref{prelim}(d), where an injector and a detector ferromagnet are situated on top of a non-magnetic channel. When electrons flow through the injector magnet to the ground contact of the channel lying below the magnet, a large number of spins oriented in the same direction as the magnetization of the injector magnet are accumulated in the channel region underneath the magnet. The gradient of this spin potential difference between the two spin orientations causes one type of spin to flow along the channel, thereby exerting non-local spin-torque on the detector magnet. The magnitude of injected spin current decays exponentially with distance between the two ferromagnets due to spin-flip processes. Apart from choosing appropriate materials with longer spin-flip lengths \cite{ji2004spin,fukuma2011giant}, a tunneling barrier can be inserted between the magnet and channel to achieve better spin injection \cite{fukuma2011giant}. Recent experiments have demonstrated non-local spin-torque induced magnetization reversal in Py/Au nanopillars located on top of a Cu wire \cite{yang2008giant}.

\subsection{Towards More Efficient Devices}

Improving the efficiency of operation of spin devices, and notably the ``write" and ``read" mechanisms is key to achieving scalable, compact and low-power neuromimetic devices. Using PMA materials is one possible alternative to reduce the critical switching current density for magnetization reversal \cite{liu2012spin,yu2014switching,liu2012current} or DW displacement \cite{ryu2013chiral,ryu2014chiral,emori2014spin,emori2013current}. Other physical mechanisms like voltage-controlled magnetic anisotropy \cite{amiri2012voltage}, magnetoelectric effect \cite{heron2011electric,franke2015reversible} or topological insulator induced spin current generation \cite{mellnik2014spin,fan2014magnetization} are also under exploration that can potentially serve as replacements for HM induced magnetization switching. Innovations in the material stack, for instance using Heusler alloys \cite{hirohata2016heusler} or anti-ferromagnetic materials \cite{yang2015domain,shiino2016antiferromagnetic} may lead to further energy benefits. Multi-level information encoded by DW position in magnets can be also potentially replaced by current induced skyrmion displacement \cite{woo2016observation,kang2016skyrmion}. While the discussion in this article will be mainly based on single-domain or DW motion based multi-domain devices with HM underlayers, the concepts can be easily extended to incorporate innovations in the material stack or the underlying physical mechanism utilized for switching \cite{jaiswal2017proposal,he2017tunable,huang2017magnetic,li2017magnetic}.

Additionally, improving the TMR effect is crucial to achieving more efficient synapses that can offer higher distinguishability for the scaling operation of the neuron inputs. While the theoretical limit of the AP and P resistance ratios is near 300 \cite{zhang2004large}, experiments have achieved a maximum variation of $600\%$ till date \cite{ikeda2008tunnel}. A roadmap issued by the IEEE Magnetics Society has predicted a variation of $1000\%$ in a time period of ten years \cite{hirohata2015roadmap}.

\section{Neuromorphic Computation: Preliminaries}
In this section, we will first describe the functionality of the major units of such neural computing models. We will also discuss different variants of neuron models (with varying degrees of bio-fidelity) and synaptic learning mechanisms. Relationship of such models to neuroscience mechanisms observed in the brain will be also established.

\begin{figure*}
\centering
\includegraphics[width = 7.1in ]{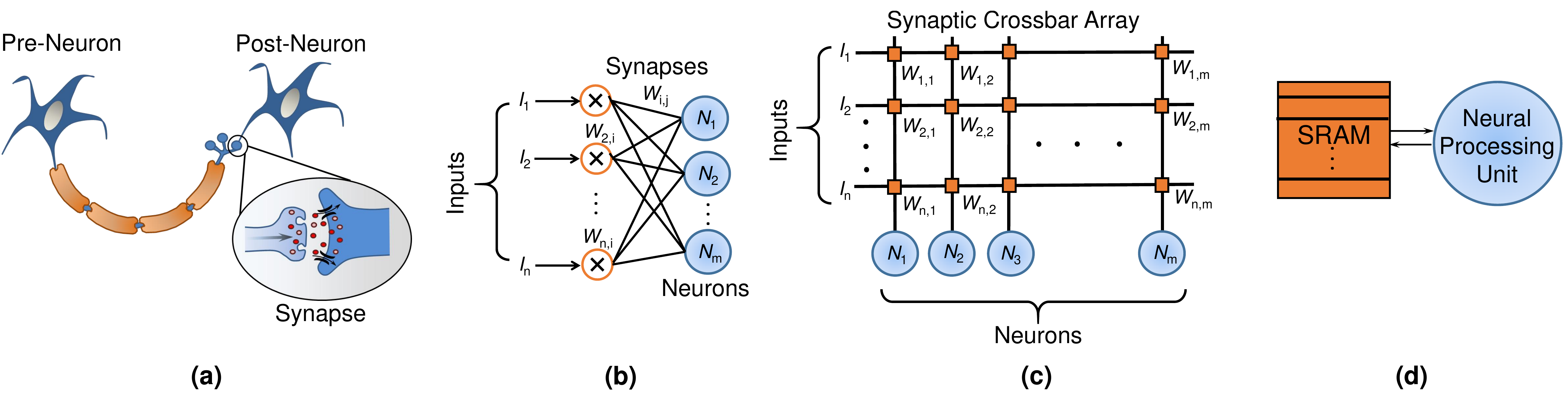}
\caption{(a) A pre-neuron transmits signals to a post-neuron through a synaptic junction. (b) Computation in a particular layer of a fully connected network can be mapped to a parallel dot-product operation between the inputs and the synaptic weights followed by neural processing for each neuron in the layer. (c) Such a computing kernel can be implemented in a crossbar array structure where programmable resistive devices encoding the synaptic weight are present at each cross-point. Input voltages applied along the rows get weighted by the synaptic conductance and provide the resultant input current (dot-product of applied voltages and synaptic conductances) to the neuron for processing. (d) In contrast, a CMOS architecture would consist of an SRAM module for synaptic weight storage. Memory access and memory leakage due to data transfer between the SRAM module and the computation core (Neural Processing Unit) constitute a significant portion of the total energy consumption.}
\label{fig1}
\end{figure*}
The main functional units of such neuromimetic computations are the neuron and the synapse. Synapses are adaptive or plastic junctions between neurons that modulate the strength of the signal being transmitted from the pre-neuron to the receiving or post-neuron. Computational tasks like pattern recognition are therefore performed by virtue of plasticity of the synapses in response to signals being transmitted between the neurons since they encode the importance level of different inputs being received by a particular neuron. Fig.\ref{fig1}(a) depicts a particular synaptic connection between a pre- and a post-neuron. Neuromorphic computation relies on the abstraction of the plasticity of the synaptic junction (governed by neuro-transmitter release at the synapse due to the incoming action potential from the pre-neuron) and the neuroscience mechanisms occurring in the post-neuron (to generate an outgoing signal to the next layer of neurons).
\subsection{Neural Computation}
Each neural computing unit receives a set of inputs from other pre-neurons through synaptic junctions. The weighted contribution from all the neurons is then summed up and processed by the neurons. The bio-fidelity level at which the ``artificial'' neuron is modelled has gradually evolved over the last few years from simple perceptrons to more biologically realistic spiking neurons \cite{de2001artificial}. Irrespective of the details of the neural model, it is worth noting the nature of neuromorphic computation being realized in such networks. Considering a set of neurons in a particular layer receiving a set of inputs through synaptic weights, the computation can be mapped to a parallel dot-product operation between the inputs and synaptic weights followed by neural processing for each neuron in the layer (Fig. \ref{fig1}(b)). Such a computing kernel is inherently suited for ``in-memory" computing platforms based on crossbar arrays of memristive devices as shown in Fig. \ref{fig1}(c) \cite{yang2013memristive, prezioso2015training}. A memristor is a nanoscale non-volatile programmable resistor. Input voltages drive the rows of the crossbar array where a resistive device encoding the synaptic weight is present at each cross-point joining a particular input to the corresponding neuron. The current flowing through a particular memristive synapse is scaled by the device conductance (synaptic scaling operation) and all such currents gets summed up along the column of the array, according to Kirchhoff’s law, and passes as the resultant input to the neuron. Additionally, due to non-volatility of the crossbar memristive elements, such architectures do not suffer from leakage concerns. In contrast, digital CMOS implementations like the IBM TrueNorth involves an architecture depicted in Fig. \ref{fig1}(d), where synaptic weights would be fetched from a Static-Random-Access-Memory (SRAM) bank to the neuron computing core \cite{merolla2014million,akopyan2015truenorth}. The inefficiency of such architectures results from the memory access and leakage energies (which usually constitutes $\sim 60-80\%$ of the total energy consumption in typical pattern recognition workloads) and the overall system performance is memory bandwidth limited. 

Let us now describe the details of neural processing across different generations. Perceptron networks consist of neurons having ``step'' transfer function (relationship between the output and input signals), i.e. they generate a high output signal if the weighted summation of neuron input crosses a particular threshold \cite{de2001artificial}. However, since their success was limited to only a very small set of simple problems, they were replaced by the ``second'' generation of ``artificial'' neurons where the transfer function of the neuron was ``non-step'', i.e. the neuron produced an analog output in response to the input stimulus \cite{de2001artificial}. Such neurons offer high recognition accuracies in a vast category of large-scale recognition tasks and are routinely utilized today as a basic building block of deep neural networks. The scalability of such neurons to more ``difficult'' problems can be attributed to the fact that a greater degree of information can be encoded in the analog neuron output in contrast to the encoded binary information in perceptron networks. A second and equally important contributing factor is the gradient of the neuron transfer function. Backpropagation \cite{hecht1988theory}, which is the underlying algorithm for training networks of such neural units, relies on the computation of the partial derivative of the error function (difference between the network output and the desired output) with respect to the synaptic weights, which in turn, is dependent on the gradient of the neuron transfer function. Hence, while a ``non-step'' neuron transfer function offers gradient information during error backpropagation, perceptrons offer gradient information only at the threshold point. A few popular ``non-step" neuron transfer functions are the Sigmoid and Rectified Linear Unit (ReLU) functions.

A more recent paradigm shift in neural computing has been the ``spiking'' neuron model, encoding a much higher degree of bio-fidelity \cite{maass1997networks}. A principal biological information that was completely ignored in the first two neuron generations was the mode of neural communication. Biological neurons communicate with each other through binary signals or spikes \cite{maass1997networks,ghosh2009spiking}. Hence, in order to account for neuron communication by means of spikes and simultaneously overcome the bottlenecks of perceptrons (neuron providing `0' - no spike and `1' signal - spike), such ``spiking'' neurons consider the input as a time-series event instead of a single value as in previous generations. The input is usually encoded in a series of time-steps and provided to the neuron. A common form of input encoding is that of a Poisson spike train, where the probability of spike generation at a particular time-step is proportional to the value of the input. This is usually referred to as ``rate'' encoding \cite{brette2015philosophy} in literature, since the number of spikes transmitted over a given timing window is proportional to the value of the input.

The most common ``spiking'' neuron model is that of the Leaky-Integrate-Fire (LIF) neuron \cite{ghosh2009spiking}, whose temporal dynamics is given by,
\begin{equation}
\label{lif}
C_{mem} \frac{dV_{mem}}{dt}=-\frac{V_{mem}}{R_{mem}}+\sum\limits_{i} w_{i}. \delta(t-t_{f,i})
\end{equation}
where $V_{mem}$ is the membrane potential, $R_{mem}$ is the membrane resistance, $C_{mem}$ is the membrane capacitance, $w_{i}$ is the synaptic weight for the $i$-th input, and $\delta(t-t_{f,i})$ is the spiking event occurring at time-instant $t_{f,i}$. When the neuron's membrane potential $V_{mem}$ crosses the threshold $V_{th}$, the membrane potential gets reset to $V_{reset}$ and does not vary for a time duration termed as the refractory period \cite{ghosh2009spiking}. Note that more bio-plausible neural models account for the modelling of a post-synaptic current that increases every time a spike is received and then decays exponentially \cite{ghosh2009spiking}. This post-synaptic current is then integrated by the LIF neuron instead of the spikes as mentioned in Eq.\ref{lif}.

It is worth noting here that ``spiking'' neuron models are not only limited to being more biologically plausible, but offers a host of advantages from hardware implementation perspective. The foremost breakthrough has been in the arena of unsupervised adaptive local learning enabled by Spike-Timing Dependent Plasticity (STDP) which has made it possible for learning functionalities to be enabled ``on-chip''. We will discuss synaptic learning in details in the next sub-section. Additionally, since such networks are `spike' or `event driven' and can perform pattern recognition by sparse distribution of spikes, they can potentially lead to sparse, event-driven hardware that exploits power-gating functionalities \cite{merolla2014million,akopyan2015truenorth}. For instance, synaptic weights can be now fetched from the SRAM bank only upon the receipt of an input event or `spike' (unlike non-spiking nets where all the synaptic weights are required to be fetched to the computing core for each input). Asynchronous event-driven communication techniques at the architecture level like Address Event Representation (AER) are also under exploration \cite{chan2007aer,indiveri2006vlsi}. At the circuit level, an additional benefit is achieved due to the replacement of a multiplier by a multiplexer for each synaptic scaling operation. Since the inputs are binary, they do not need to be multiplied by the synaptic weights but can be transmitted to the neural computing core in case a `spike' is received \cite{han2016energy}. Note that the loss of information due to binary inputs is compensated by temporal encoding over the time-steps of the spike train. However, the advantages due to reduced power consumption of spiking networks (event-driven hardware) far outweigh the cost of increased delay for inference (temporal encoding) \cite{han2016energy}.

Due to such inherent advantages of Spiking Neural Networks (SNNs) at the hardware level, there has been significant interest in recent years to convert non-spiking nets to SNNs by replacing the original neurons by ``spiking'' neurons after training \cite{diehl2015fast}. The main motivation behind the conversion stems from the fact that while non-spiking nets can be trained with very high classification accuracies at large-scale recognition tasks using backpropagation, achieving similar accuracies in STDP trained spiking networks is still an active research area. The ``spiking'' neuron model typically used for such conversion schemes has been the Integrate-Fire (IF) model which is equivalent to the LIF neuron without any leak term in the membrane potential. Such an IF neuron without any refractory period has been shown to be a firing-rate approximation of the ReLU unit mentioned previously \cite{cao2015spiking}. This is apparent from the fact that higher the value of the input for the ReLU, higher is the value of the neuron output. Similarly, for the IF neuron, higher is the rate of input spikes, higher is the number of transmitted output spikes. 

However, note the fact that the above ``spiking'' neuron computing models are completely deterministic and do not account for the noisy probabilistic neural computation that actually occurs in the human brain. Recent proposals have investigated stochastic neural models that abstract the neural computation by a probability distribution function that varies as a function of the input being received by the neuron at each time-step of computation \cite{wallace2011emergent,benayoun2010avalanches,nessler2009stdp,nessler2013bayesian}. The variation is usually characterized by a non-linear functionality, for instance a sigmoid function. Such probabilistic neural computation has been observed in `pyramidal' spiking neurons in the cortex and recent research proposals have investigated the possibility of performing Bayesian computation in cortical microcircuits of stochastic neurons \cite{nessler2009stdp,nessler2013bayesian}. Additionally, such stochastic neural computational units have been also used in Restricted Boltzmann Machines and Deep Belief Networks \cite{hinton2006reducing} trained by Contrastive Divergence \cite{hinton2006fast}. Such probabilistic ``spiking'' neural models are particularly interesting for spintronic device applications since such devices are inherently characterized by a time-varying thermal noise leading to stochastic behavior. 

We would like to conclude this section on neural computation by a brief discussion on an additional neuroscience mechanism termed as homeostasis \cite{diehl2015unsupervised} that is also routinely utilized in SNN based pattern recognition systems. It is a spike frequency adaption mechanism wherein the neuron threshold increases by a specific amount every-time the neuron spikes. This ensures that as a neuron starts to dominate the spiking pattern in a particular pool of neurons, it also becomes progressively difficult for that particular neuron to spike in the future. We will discuss the manner in which such homeostasis effects assist in performing pattern recognition.

\subsection{Spike-Timing Dependent Plasticity}

As mentioned in the previous section, prior to the advent of SNNs, synaptic learning was achieved primarily by backpropagation algorithm \cite{hecht1988theory}. This is a supervised training algorithm where the neural network is trained with a particular set of inputs that are associated with specific class labels or categories. The algorithm aims at finding the optimal set of synaptic weights by minimizing the error function (difference between class labels and actual network outputs) using gradient descent algorithm. Readers are referred to Ref. \cite{hecht1988theory} for details on the backpropagation algorithm. A few key points worth noting is the supervised nature of the training algorithm and the synaptic weight update scheme which is not only dependent on the outputs of neurons in other layers of the network but also require a backward pass of the gradient computation through the entire network. This has broadly limited the scope of specialized hardware to implement backpropagation on-chip due to expensive power and area requirements of the underlying hardware.

The number of applications requiring some form of intelligence in present day Internet of Things (IoT) technologies like mobiles and wearables are huge and often require embedded on-chip intelligence since it is often not possible to transmit data in real-time to cloud for computing. Further, it is also not practical to have supervised learning algorithms to implement pattern recognition systems since real-time data will be mostly unlabeled (without any specific categories). Hence, unsupervised hardware-inexpensive synaptic learning mechanisms is a key requirement for the implementation of on-chip learning.
\begin{figure*}
\centering
\includegraphics[width = 7.0in ]{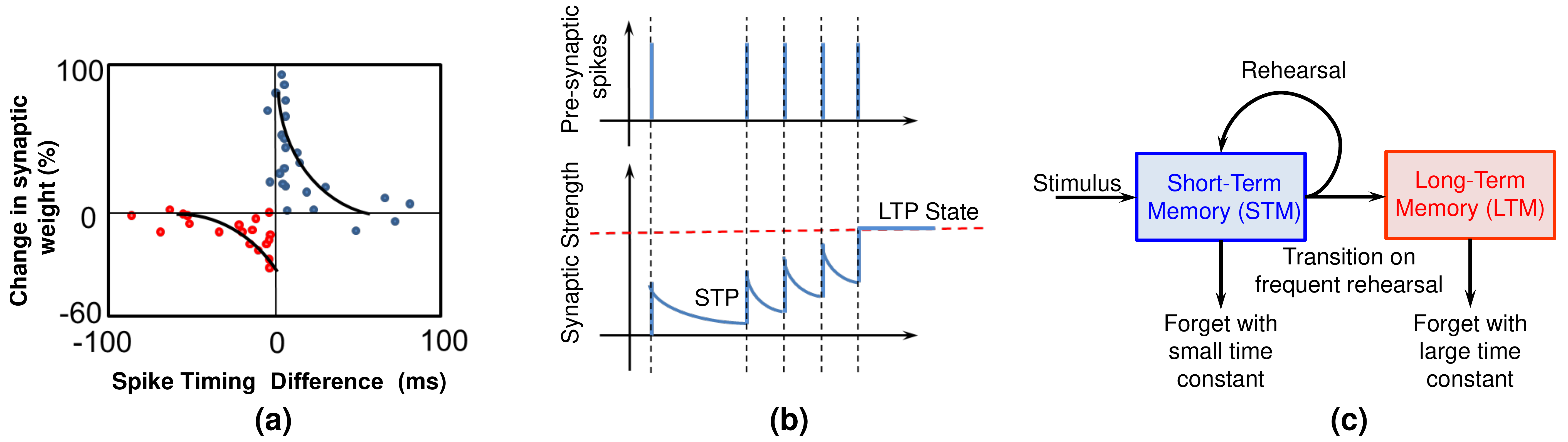}
\caption{(a) Spike-Timing Dependent Plasticity (STDP) measurements obtained from rat hippocampal glutamatergic synapses \cite{bi2001synaptic}. STDP learning rule can be formulated by considering that the synaptic weight potentiates (depresses) if the pre-neuron spikes before (after) the post-neuron. The variation is exponential with spike-timing difference. (b) The synaptic strength increases momentarily on the receipt of a pre-synaptic spike but starts decaying back to the initial value in the absence of spikes. This is referred to as Short-Term Plasticity (STP). On frequent stimulation, the synapse strengthens to a long-term stable state. This is referred to as Long-Term Potentiation (LTP). (c) STP-LTP is often correlated to the concept of Short-Term Memory (STM) and Long-Term Memory (LTM). While information is initially stored in the STM, it gets transferred to the LTM on frequent rehearsal of the input stimulus.}
\label{fig2}
\end{figure*}

A more bio-realistic and hardware-friendly approach to synaptic learning in comparison to backpropagation is the STDP learning rule in SNNs, which is based on measurements obtained from rat hippocampal glutamatergic synapses \cite{bi2001synaptic} (Fig. \ref{fig2}(a)). According to this theory, the synaptic weight is modulated depending on the spiking patterns of the pre-neuron and post-neuron. The synaptic weight increases (decreases) if the pre-neuron spikes before (after) the post-neuron. Intuitively, this signifies that the synapse strength should increase if the pre-neuron spikes before the post-neuron as the pre-neuron and post-neuron appear to be temporally correlated. The relative change in synaptic strength decreases exponentially with the timing difference between the pre-neuron and post-neuron spikes. The STDP characteristics can be formulated in a mathematical framework as follows, 
\begin{equation}
\begin{aligned}
\label{stdp}
\Delta w &=A_{+}\exp\left(\frac{-\Delta t}{\tau_{+}}\right), \Delta t > 0\\
&=-A_{-}\exp\left(\frac{\Delta t}{\tau_{-}}\right), \Delta t < 0
\end{aligned}
\end{equation}
Here, $A_{+}, A_{-}, \tau_{+}$ and $\tau_{-}$ are constants and $\Delta t = t_{post}-t_{pre}$, where $t_{pre}$ and $t_{post}$ are the time-instants of pre- and post-synaptic firings respectively. We will refer to the case of $\Delta t >0$ ($\Delta t <0$) as the positive (negative) time window for learning for the rest of this text. Note that this learning mechanism is unsupervised since no prior information about input class or label is necessary. Further, synaptic weight update is completely local since it is modulated depending on the activities of only the neurons it connects. This has enabled learning functionalities to be implemented on-chip at much lower hardware costs. Although pattern recognition systems with high accuracies based on STDP learning are still in preliminary stage, competitive accuracies in typical digit recognition and sparse encoding workloads have been already achieved \cite{diehl2015unsupervised}. Note that the above STDP learning rule is referred to as anti-symmetric STDP and has been the most popular learning mechanism for training SNNs. However, other variants of STDP have been also observed in neuroscience studies and have been utilized in different genres of recognition tasks \cite{kuzum2012low}. 

We will discuss STDP implementation in spintronic synapses in later sections. However, a primary concern for such spintronic synapses, and in general any memristive synapse technology, is the bit resolution at aggressively scaled device dimensions. Driven by this fact, researchers have proposed variants of STDP learning based on single-bit synapses \cite{modha2011stochastic,vincent2015spin,srinivasan2016magnetic} where the multi-bit requirement is replaced by probabilistic synaptic weight update. It has been already mentioned that spintronic devices exhibit an inherent stochasticity during the switching process which has been mainly attributed to the time-varying thermal noise \cite{scholz2001micromagnetic}. Hence, the STDP framework described in Eq. \ref{stdp} can be modified in this scenario as the probability of binary synaptic state change (instead of analog weight change) to offer a direct correspondence to stochastic switching behavior of single-bit nanoelectronic synapses. Stochastic single-bit synaptic learning achieving competitive accuracies in digit recognition applications has been recently demonstrated in SNNs \cite{srinivasan2016magnetic}. 
\subsection{Volatile Synaptic Learning}

The exact mechanisms that underlie learning or plasticity of synapses is highly debated and still unknown. While STDP has been a popular viewpoint of explaining synaptic plasticity, there has been some research studies that attempt to explain synaptic plasticity from an alternative volatile learning plasticity viewpoint. This is referred to in literature as Short-Term Plasticity (STP) and Long-Term Potentiation (LTP) \cite{zucker2002short,martin2000synaptic}. The theory postulates that synapses undergo inherent volatile state changes upon receipt of incoming action potentials (due to release of neurotransmitters). In case the action potentials are received infrequently, the neurotransmitter concentration decays to the background value after the action potential is removed and hence the synaptic plasticity remains unchanged (STP). However, as more frequent action potentials are received, the ionic neurotransmitter concentration starts increasing and ultimately the synapse switches to a stable long-term state (LTP). Hence, while STDP is a form of non-volatile synaptic learning, STP-LTP models synaptic plasticity as a form of frequency-dependent volatile synaptic learning. While adoption of STP and LTP concepts in SNNs for usage in pattern recognition is still an area of active research, it offers the promise of adaptive learning where the network might be able to unlearn itself in response to changing environments, which might not be possible to achieve by non-volatile STDP learning rule. 

Such a learning mechanism is in accordance to the volatile forgetting nature of human memory and has been often correlated to Short-Term Memory (STM) and Long-Term Memory (LTM) psychological models proposed by Atkinson and Shiffrin \cite{atkinson1968human,lamprecht2004structural}. The model is equivalent to STP and LTP where the synaptic element can be viewed to be analogous to human memory. Input information is received and stored in the STM and only gets transferred to LTM if the input is received with sufficient frequency. The characteristic difference between STM and LTM is that while information is stored for a limited period in STM (analogous to volatile meta-stable synaptic state change in response to input stimulus), LTM retains the information for a much longer period of time (analogous to long-term stable synaptic state). Fig. \ref{fig2}(b) and (c) illustrates the concepts of STP-LTP and STM-LTM respectively. It is worth noting here that psychological STM-LTM concepts have been also harnessed to model the computational units of Recurrent Neural Network (RNN) architectures \cite{hochreiter1997long}.
\subsection{Network Connectivity}
\begin{figure*}
\centering
\includegraphics[width = 6.5in ]{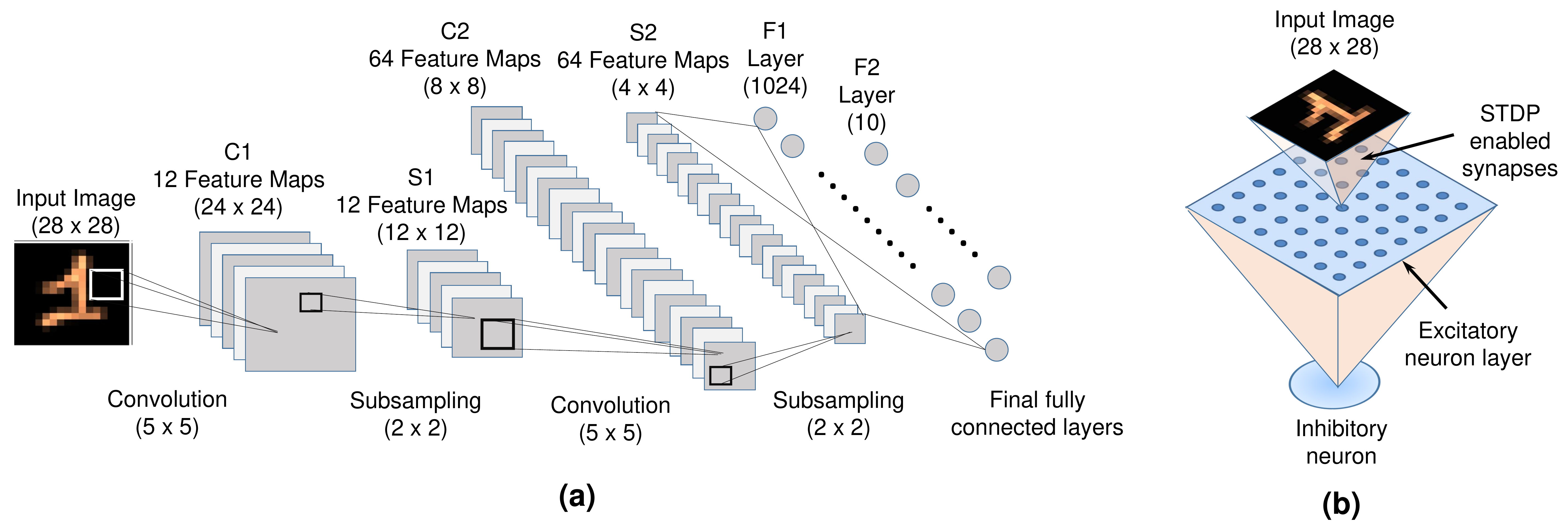}
\caption{(a) A Deep Convolutional Neural Network (CNN) consists of alternate cascaded layers of convolution and subsampling terminated by a fully connected output layer. The figure depicts a typical CNN network used for digit recognition (28x28-12c5-2s-64c5-2s-10o). (b) A network typically used for studying STDP is shown. Such connections have been observed in cortical microcircuits of pyramidal neurons in the brain. It consists of an excitatory layer of neurons that receives spike trains from the input in an all-to-all fashion. Lateral inhibition and homeostasis promotes STDP learning in such single layer networks.}
\label{fig5}
\end{figure*}
The discussion so far has been limited broadly to the functionalities exhibited by the fundamental units in neuromorphic systems. However, in order to construct pattern recognition systems based on these units, specific network connections and topologies are necessary. Initial studies in neural networks mainly focused on fully-connected nets (FCNs), where neurons are arranged in different layers and connected in an all-to-all fashion, as shown in Fig. \ref{fig1}(b). However, such simple network connectivity failed to be invariant to translation or scaling of input patterns. Further, FCNs with larger number of neurons/layers implies storage of a huge set of synaptic weights along with higher degree of neuron connectivity between layers which limits its scalability to large-scale cognitive tasks.

Deep networks based on convolution operations have been able to overcome most of these challenges. The inspiration behind such a connectivity is based on the seminal work of Hubel and Wiesel which revealed that the animal cortex consists of cells which are sensitive to specific areas of the entire visual field (implying a local connectivity for each neuron) and that they function as filters for that particular receptive field \cite{hubel1968receptive}. Further, a certain category of cells were found to be sensitive to edge-like features in the visual field while another category of cells were found to be invariant to the location of the pattern in the receptive field \cite{hubel1968receptive}. Such mechanisms served as the main motivation behind the structure of Convolutional Neural Networks (CNNs).

Fig. \ref{fig5}(a) shows the CNN structure. Drawing inspiration from the hierarchical arrangement of layers in the visual cortex, CNNs consist of a number of cascaded stages where each stage consists of a convolution layer (C) followed by a sub-sampling layer (S). Each C layer is characterized by a set of trained weight kernels that is used to convolve with the input maps for that particular layer. For instance, in an image recognition system the input map for the first layer of a network would be the entire image being classified. Each kernel is then convolved with the entire image to produce an equivalent number of output maps. Each neuron in the output map therefore has limited connectivity (equal to the size of the convolution kernel). Additionally, the network offers resiliency to image translation and scaling due to the convolution operation. The C stage is usually followed by an S layer which performs an averaging operation over non-overlapping subsampling windows of each output map to reduce their dimensionality. As the depth of the layer increases, the number of maps increases with decreasing dimensionality. Ultimately the final two layers are usually fully connected and the number of neurons in the output layer equals the number of classes in the recognition problem. Due to the limited fan-in of each neuron, sparse neural connectivity is achieved. Additionally number of synaptic weights to be learnt during training is also reduced, due to the shared weight kernel being convolved across the entire map, thereby resulting in significantly reduced training time. 

An alternative network architecture that has been popular in the domain of STDP learning enabled SNNs has been shown in Fig. \ref{fig5}(b) \cite{diehl2015unsupervised}. Such connections are again inspired from cortical microcircuits of pyramidal neurons observed in the brain. The network consists of a layer of neurons that receive input spike trains through excitatory (positive) synaptic weights in an all-to-all fashion. The network is also associated with a lateral inhibitory signal that triggers a negative spike signal whenever one of the neurons in the layer spikes. In order to prevent single neurons from dominating the spiking pattern due to lateral inhibition, the ``spiking" neurons are enabled with homeostasis functionality. STDP in the excitatory synaptic connections in such networks can assist each neuron to respond selectively to specific classes of input patterns. Note that training deeper networks enabled by STDP is still an area of active research.

While the discussion in this section mainly focused on feedforward networks without any directed loops, RNN architectures are also becoming increasingly popular for sequence learning tasks like language modelling \cite{mikolov2010recurrent}, handwriting prediction and generation \cite{graves2013generating}, speech recognition \cite{graves2013speech}, among others. The only difference between RNNs and standard feedforward networks is the fact that the computational units or neurons receive its own output from the previous time-step as its input in the current time-step (in addition to external inputs). Such a memory effect in RNNs enables it to perform context learning in sequential inputs. However, note that the main functionalities of the computational units -- the neurons and synapses remains unaltered, thereby allowing the same synaptic/neural spin-devices to be used in these different algorithmic architectures. This will be discussed in details in the next section.

\section{Spintronic Device Proposals and Correspondence to Neural and Synaptic Functionalities}
Nanoscale programmable resistive devices mimicking neural and synaptic functionalities is imperative towards the realization of energy-efficient neuromorphic architectures. The field of neuromorphic computing, wherein research effort is directed to mimic neural and synaptic mechanisms by the underlying device physics, was pioneered by Carver Mead in the 1980s \cite{mead1990neuromorphic}. He proposed that CMOS transistors operating in subthreshold region can be utilized to implement neuromimetic computations since the main mechanism of carrier transport in that operating regime is diffusion, thereby emulating the mechanism of ion flow in biological neuron channels \cite{mead1990neuromorphic}. Although such sub-threshold CMOS neuron and synapse designs are still being investigated by various research groups \cite{chicca2014neuromorphic}, they require multiple transistors and feedback mechanisms to mimic the functionality of neurons/synapses. The first work on spintronic neuromorphic computing can be traced back to the work of Krysteczko $et$ $al.$ where they explored the possibility of implementing memristive functionalities in MTJ structures through voltage induced switching phenomena \cite{krzysteczko2012memristive}. 

\subsection{Spin-Torque Neuristors}
\begin{figure*}
\centering
\includegraphics[width = 4.0in]{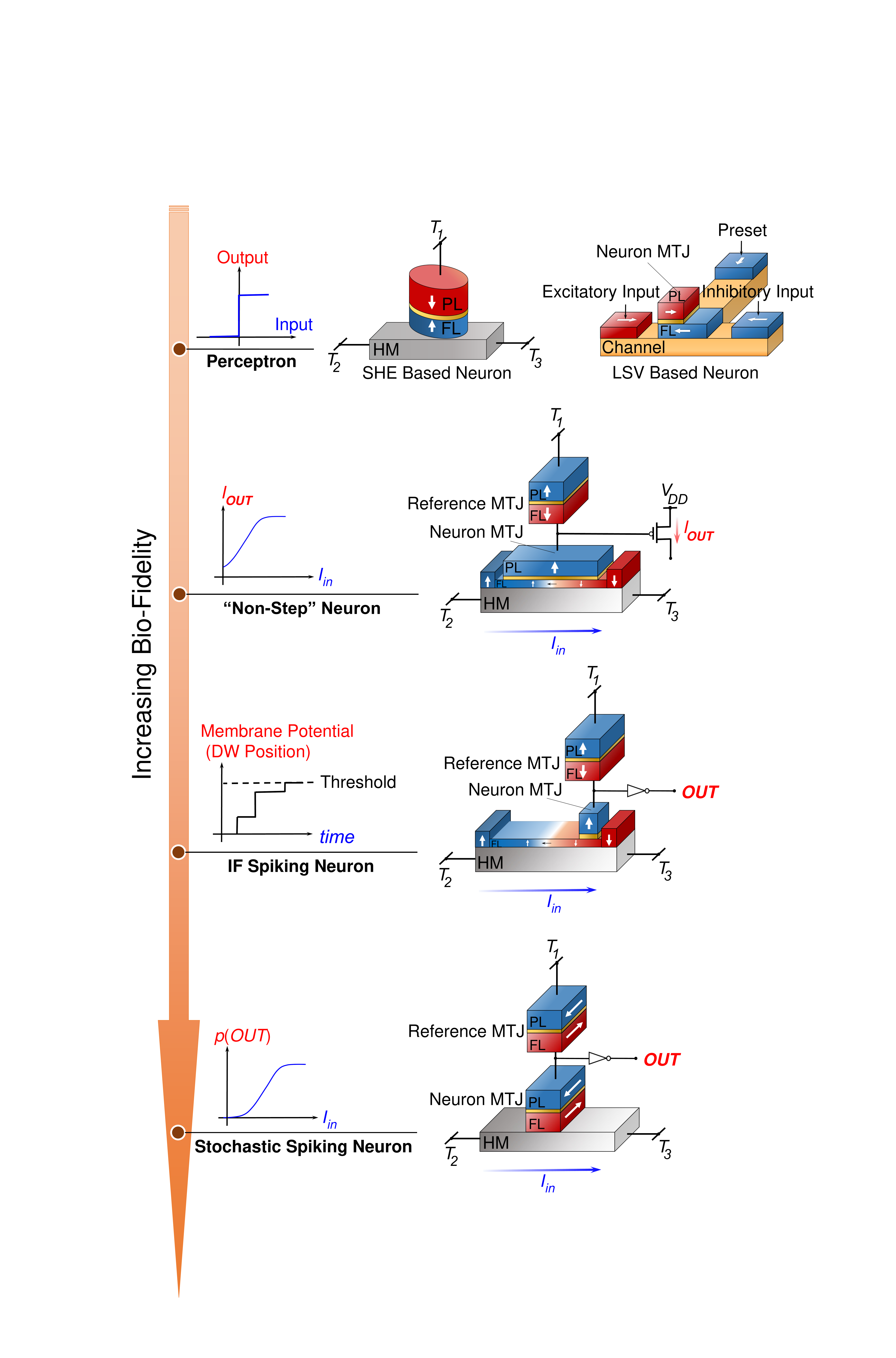}
\caption{Spin-torque neuristors with different degrees of bio-fidelity are shown. Perceptron or ``step" neurons can be implemented in SHE based neuron structures where a current flowing through an underlying HM layer orients a PMA magnet lying on top along the unstable ``hard-axis". Subsequently the direction of current flowing through the PL orients the magnet to either of the stable ``easy-axis" directions. A complementary device structure can be envisioned using the LSV concept by injecting spins oriented along the ``hard-axis" in a non-magnetic channel using a ``Preset" magnet. ``Non-step" neuron functionalities can be implemented in DW motion based device structures by interfacing the Neuron MTJ with a Reference MTJ. A similar device structure with the MTJ located at the edge of the FL can be used to implement an IF ``spiking" neuron. Stochastic ``spiking" neuron functionalities can be implemented in mono-domain neural device structures by exploiting the underlying probabilistic MTJ magnetization dynamics. }
\label{neuron}
\end{figure*}

In this section, we will review different spintronic device structure proposals that can potentially offer a direct correspondence to neuronal computations with varying degrees of bio-fidelity. Fig. \ref{neuron} depicts various spintronic devices mimicking neurons of different computing generations from ``step" to ``spiking" neurons. We will begin our discussion on the neuronal devices by considering it receives a resultant weighted synaptic current input. Synaptic device structures and interfacing of synaptic arrays with neuronal devices for generating the input synaptic current will be discussed in the next sections.

\subsubsection{Step Neurons}
Let us begin this section by noting the functional similarity between a ``step" neuron transfer function and a mono-domain MTJ switching event. The MTJ switches between the two stable P and AP states provided the switching current magnitude is greater than a particular threshold. Consequently, in order to emulate the ``step" neuron functionality with neuron threshold at the origin, the input current to an MTJ neuron has to be greater than the switching current requirement, which in turn, increases the operating voltage of the MTJ. Ref. \cite{2015spinijcnn} investigated the design of an MTJ based neuron for the implementation of a ``step"-transfer function neural network. In order to reduce the input synaptic current magnitude, the MTJ was initialized to the AP state and provided with a bias current that was equal to the critical current requirement for MTJ switching to the P state. Hence, a small magnitude of synaptic current (positive or negative) would ensure MTJ switching to either the P state or remaining in the original AP state. However, due to the high bias and reset current requirements, energy improvements for such MTJ ``step"-neuronal devices was highly limited and Ref. \cite{2015spinijcnn} demonstrated that $\sim 1.63\times$ energy improvement could be achieved by such neurons in comparison to a digital CMOS neuron implementation. Note that in this work, the focus point has been the mapping of simply the MTJ switching event to a neuron functionality while the internal time-domain magnetization dynamics has not been considered. As we will show later, utilization of the stochastic MTJ switching dynamics due to time-varying thermal noise to model neural computations can lead to ``spiking" neuron implementations with higher bio-fidelity and enhanced recognition performances for computing platforms. 

Continuing our discussion on simply the MTJ switching event to mimic a ``step" neuron, the energy consumption can be drastically reduced in case the MTJ is initialized to an unstable magnetization state prior to the switching process. This would assist in reducing the critical current requirement for the switching process, since a very small magnitude of input synaptic current can now enable the switching process to either of the two stable states (depending on the input spin current direction) by overcoming thermal fluctuations. This concept was utilized in Ref. \cite{2015spinapl} in a spintronic device structure where a PMA magnet lies on top of a HM and is operated in two subsequent stages of ``Preset" and ``Evaluation". Note that in Section II.3, we mentioned that PMA magnets cannot be switched solely in presence of SOT since in-plane spins are injected by current flowing through the underlying HM into the PMA nanomagnet lying on top. Given sufficient magnitude of this ``Preset" current flowing through the HM, the in-plane injected spins would be able to orient the magnetization along the ``hard-axis", i.e., the in-plane direction. If the ``Preset" stage is followed by an ``Evaluation" stage (during the relaxation time-constant of the magnet: time duration during which the magnet does not relax to either of the two ``easy"-axis directions in presence of thermal noise once the ``Preset" pulse is removed) where input synaptic current flows through the MTJ, then its direction determines the final state of the MTJ. Due to the initialization of the magnet along an unstable axis, deterministic MTJ switching can be now achieved at much lower current magnitudes. Studies performed for a PMA CoFeB magnet of dimensions $\frac{\pi}{4} \times 40 \times 40 \times 1.5 nm^3$ and magnetic parameters based on measurements reported in Ref. \cite{pai2012spin} for $\beta$-W underlayer demonstrated that the neuron operation involved $\sim 15fJ$ energy consumption, thereby leading to $\sim 3\times$ improvement in energy consumption over a corresponding CMOS implementation.

An alternative approach utilizing the LSV concept introduced in Section II.4 for spin-transport was explored by Sharad $et$ $al.$ in Ref. \cite{sharad2012spin}. The device operation is demonstrated in Fig. \ref{neuron} where the main neuron magnet is interfaced with magnets (Excitatory, Inhibitory, Preset) on top of a non-magnetic channel. The ``Preset" magnet has its ``easy"-axis orthogonal to that of the other magnets. Hence, during the ``Preset" phase, charge current pulse through it causes the neuron magnet to be initialized along the unstable ``hard"-axis due to spin transport through the non-magnetic channel. Subsequently, during the ``Evaluation" stage, the positive synaptic inputs drive the excitatory magnet while negative synaptic inputs drive the inhibitory magnet. These magnets have their magnetization directions AP to each other. Hence, the resultant synaptic current injected to the neuron magnet determines its final orientation state. A limiting factor to the scalability of this approach to large neural network designs is the spin-flip process which causes the spin-current to decay exponentially in the channel. This, in turn, constrains the fan-in capability of each neuron. Hence proper channel materials for efficient spin current injection are under exploration. Additionally, in order to increase spin-current injection at the magnet-channel interface, a tunneling barrier like MgO is utilized between the two materials, thereby increasing the energy consumption for charge current being injected through the synapses. Alternative device proposals utilizing spin-transfer torque oscillators for implementing ``step" neuron functionalities have been also proposed in literature \cite{yogendra2015coupled}.

\subsubsection{Non-Step Neurons}
Let us now proceed to the implementation of ``non-step" neuron functionalities in spintronic devices. Note that since an MTJ with a mono-domain FL consists of two stable states, only two distinct neuron outputs can be represented by such a device structure. However, for a multi-domain FL, where the magnet consists of two oppositely polarized magnetization regions separated by the DW, the device can exhibit multi-resistive states \cite{fan2015stt}. Experimentally first demonstrated by Chanthbouala $et$ $al.$ as a three-level programmable resistor \cite{chanthbouala2011vertical}, a multi-level DW motion based resistive device was recently shown to exhibit 15-20 intermediate resistive states \cite{lequeux2016magnetic}. 

As shown in Fig. \ref{neuron}, the device structure consists of an MTJ structure where the FL is a DWM lying on top of a HM layer (for energy efficiency) \cite{sengupta2016proposal,sengupta_allspinsnn}. The underlying device physics for transverse N\'{e}el DW motion in such PMA magnetic multilayers due to charge current flow through the HM has been discussed in Section II. Note that a complementary device structure utilizing spin-orbit torque induced Bloch DW motion was also investigated in Ref. \cite{sengupta2015spin}. Although the discussion henceforth will be based on N\'{e}el wall motion, the concepts are equally valid for device structures utilizing Bloch DW motion. The FL is surrounded by two PLs on either side to ensure that the DW stabilizes at the opposite edges of the FL for large magnitudes of the current flowing through the underlying HM. 

The operation of such a multi-terminal device occurs in two subsequent ``write" and ``read" stages. During the ``write" stage, the magnitude of current flowing through the HM (``write" current) programs the position of the DW in the FL of the MTJ structure. The DW displacement increases linearly with the magnitude of the input synaptic current flowing through the underlayer ($I_{in}$) between terminals $T_2$ and $T_3$. After the ``write" phase, terminal $T_1$ is activated instead of $T_2$ which enables the ``read" current path in the device between terminals $T_1$ and $T_3$. Such decoupled ``read" and ``write" current paths not only assist in optimizing the ``write" and ``read" peripheral circuits independently but also enable a low value of resistance in the path of the ``write" current (mainly the resistance of the underlying HM layer). As we will discuss in the next section, a crucial functionality that is required for nanoelectronic neurons is low input ``write" resistance for proper operation of neuromorphic crossbar arrays. It is the decoupled nature of the ``write" and ``read" current paths of such multi-terminal devices that have made it possible for spintronic devices to be utilized not only as a synapse, but also as a neuron.

The DW position of the FL is sensed by a simple resistive divider, as shown in Fig. \ref{neuron}, where the neuronal device is interfaced with a Reference MTJ which is always fixed to the AP state. The ``read" current can be maintained to sufficiently low magnitudes by ensuring proper oxide thickness of the neuronal and Reference MTJs which assists in achieving ``disturb-free read" of the neuron MTJ. This resistive divider drives a transistor operating in saturation regime (in order to ensure that the supplied current to the fan-out resistive synapses is independent of the magnitude of the interfaced synaptic resistances). As the magnitude of the input current $I_{in}$ increases, the resistance of the neuronal device reduces due to decrease in the proportion of the AP domain in the MTJ device. This, in turn, causes the current provided by the output transistor ($I_{out}$) to increase. It can be shown that the transfer function (relationship between $I_{out}$ and $I_{in}$) of such a device is approximately linear by performing a device-circuit co-design \cite{sengupta2016proposal}. After every ``read" cycle, the neuron is ``reset" for the next operation by passing a current through the HM in the opposite direction to initialize the DW at the opposite edge of the MTJ. Micromagnetic simulations based on typical device parameters obtained experimentally from magnetometric measurements of Ta(3nm)/Pt(3nm)/CoFe(0.6nm)/MgO(1.8nm)/Ta(2nm) nanostrips \cite{emori2014spin} demonstrate that the DW can be completely displaced from one edge of a FL (dimension: $80 nm \times 20 nm$) to the other by $10.6\mu A$ charge current in a duration of $2ns$, thereby resulting in a total ``write" and ``reset" energy consumption of $0.1fJ$. Such energy-efficient SHE induced DW motion in magnetic multilayer devices can potentially lead to neuronal device structures that would be able to achieve multi-level neuronal states and thereby provide improved cognitive functionalities.

\subsubsection{Spiking Neurons}
Let us begin the discussion on ``spiking" neurons by noting the similarity between the current integrating property of DW motion and the functionality of an IF ``spiking" neuron. Considering input spikes (current pulses) flowing through the HM layer of an FM-HM bilayer structure at different time-steps, the DW would be displaced by an amount proportional to the magnitude of the input current pulse at each time-step whenever a spike is received. The IF functionality can be easily implemented in a slightly modified device structure, shown in Fig. \ref{neuron}, where the MTJ is located at the extreme edge of the FL and triggers an output spike (high voltage level at the output inverter) corresponding to the time-step when the DW reaches the other edge of the FL (analogous to neuron membrane potential crossing a particular threshold) \cite{sengupta_allspinsnn}. The leak functionality can be implemented by passing a current through the HM in the opposite direction at every time-step.

As mentioned previously, multi-level neuron states provided by DW motion based spintronic devices can be replaced by binary neuron states obtained from single-domain MTJ structures in case the time-domain magnetization variation of the magnet is considered. The magnetization dynamics of a nano-magnet described by Eq. ~\ref{llg} can be reformulated by simple algebraic manipulations as,
\begin{equation}
\label{llgm}
\begin{aligned}
\frac{1+\alpha^2}{\gamma} \frac {d\widehat {\textbf {m}}} {dt} =-&(\widehat {\textbf {m}} \times \textbf {H}_{eff})- \alpha (\widehat {\textbf {m}} \times \widehat {\textbf {m}} \times \textbf {H}_{eff})\\
&+\frac{1}{q\gamma N_{s}} (\alpha(\widehat {\textbf {m}} \times \textbf {I}_s )-(\widehat {\textbf {m}} \times \widehat {\textbf {m}} \times \textbf {I}_s) )
\end{aligned}
\end{equation}
Considering the device magnetization to represent the neuron membrane potential, the above equation bears resemblance to LIF characteristics of a ``spiking" neuron described in Eq. \ref{lif}. The first two terms on the RHS of Eq. \ref{llgm} represent the leak term in the magnetization state while the last term denotes the integrating term for an input spin current stimuli. Hence, in the presence of an input spike (current pulse), the magnetization starts integrating (switching) towards the opposite stable magnetization state. However, in case the pulse is removed before the entire switching event can take place, the magnetization starts leaking back toward the original magnetization state. In order to reduce the critical switching current requirement and to reduce the input ``write" resistance of the neuron, we will consider SHE-induced MTJ switching due to charge current flow through an underlying HM layer (Fig. \ref{neuron}). 

%

Once the magnet switches to the opposite magnetization state, the neuron has to be ``reset" due to the occurence of the ``firing" event. Hence, in order to sense the neuron state, the device is required to be operated in successive ``read" and ``write" cycles. Each ``write" cycle can correspond to a particular time-step of operation of the spiking network. The neuron receives weighted summation of the spike currents as its input. Since the magnetization dynamics of the MTJ is characterized by thermal noise at non-zero temperatures (in addition to the LIF characteristics discussed previously), the MTJ neuron functionality can be formulated as a stochastic ``spiking" neuron. As shown in Fig. \ref{neuron}, the MTJ exhibits non-linear stochastic switching or firing characteristics in response to the magnitude of the input current stimuli provided in a particular ``write" cycle (time-step). Unsupervised \cite{sengupta2015magnetic}/ supervised \cite{sengupta2016probabilistic} networks enabled by such probabilistic neurons will be discussed in later sections. The ``write" cycle is followed by a ``read" stage to determine the MTJ resistance (using the resistive divider driving an inverter described previously). The MTJ is ``reset" in case a spike is generated. The average neuronal energy consumption determined for the input current ($\sim 71\mu A$) necessary to switch an elliptic IMA magnet of dimensions $\frac{\pi}{4} \times 100 \times 40 nm^2$ for CoFe($1.2nm$)-W($2nm$) MTJ with a probability of 0.5 is evaluated to be $\sim 1fJ$ for a ``write" cycle duration of $0.5ns$ \cite{sengupta2015magnetic}. It is worth noting here that alternative stochastic neural computing models enabled by other post-CMOS technologies like phase change memories have been also explored in literature \cite{tuma2016stochastic}.

\subsubsection{Emergent Neural Behavior: Bursting, Oscillations, Stochastic Resonance}

While the modeling of neural spiking behavior has received a lot of attention from the neuromorphic community, not much is known about the usage of other neural phenomena that can be potentially used for cognitive tasks. Nevertheless, hardware emulation of such emerging neural behavior is still important from brain modeling and simulation perspective. For instance, authors in Ref. \cite{krzysteczko2012memristive} use the MTJ in a ``back-hopping" regime where the MTJ is initialized in an unstable state. Subsequently the MTJ resistance switches back and forth between a stable and the unstable state and after some time duration settles in the stable state. The authors infer from this random telegraphic switching characteristics that it resembles spike bursts that are emitted by a biological neuron, also referred to as ``bursting" behavior.

It has been also observed that networks of neurons in the brain oscillate in synchrony in response to specific signals. State-locking of multiple spin-torque oscillators through various injection, field or spin-wave locking mechanisms \cite{kaka2005mutual,pufall2015physical} can be exploited to model such oscillatory phenomena. Recently synchronization of nine spin-Hall nano-oscillators have been demonstrated experimentally \cite{awad2016long}. Recently neuromorphic computing platforms utilizing spintronic oscillators have been demonstrated \cite{torrejon2017neuromorphic}. Additionally spin-torque devices, being inherently thresholding devices characterized by stochasticity, can be used to implement stochastic resonance \cite{cheng2010nonadiabatic} (a phenomena used for improving the signal-to-noise-ratio due to the addition of an optimum level of noise that causes the biological system under consideration to resonate \cite{moss2004stochastic,douglass1993noise}).

We would like to conclude this section by noting the two main device structures that will be used for the rest of this discussion - the DW motion based bilayer structure used as a ``non-step"/IF ``spiking" neuron and the single-domain MTJ based device used as a stochastic ``spiking" neuron. These devices will be used to implement deterministic/probabilistic STDP in multi-/single-bit synapses respectively in the next section.

\subsection{Spin-Torque Synapses}
\subsubsection{Spike-Timing Dependent Plasticity}
The mechanism that lends cognitive capabilities to networks of interconnected neurons is the plasticity of the synaptic junctions. For a vast majority of these plasticity mechanisms, the synaptic conductance is modulated depending on the time-difference between the spikes of the neurons it connects. Let us first consider the implementation of STDP in the DW motion based device structure introduced in the previous section. The device conductance between terminals $T_{1}$ and $T_3$ is dominated by the MTJ conductance which varies linearly with the domain wall position. Let us denote the conductance of the device when the FM magnetization is P (AP) to the PL as $G_{P} (G_{AP})$, i.e. the domain wall is at the extreme right (left) of the FM. Thus, for an intermediate position of the domain wall at a location $x$ from the left-edge of the MTJ, the device conductance between terminals $T_1$ and $T_3$ is given by,
\begin{equation}
G_{eq} = G_{P}.\frac{x}{L} + G_{AP}. \left(1 - \frac{x}{L}\right) + G_{DW} 
\label{eq:res_model}
\end{equation}
where $L$ denotes the length of the MTJ excluding the domain wall width and $G_{DW}$ represents the conductance of the wall region. For a given time duration, it can be shown from micromagnetic simulations that the programming current magnitude, $J$, is directly proportional to the DW displacement, $\Delta x$ \cite{sengupta2016proposal,sengupta_allspinsnn,sengupta2016vision}. Since, $\Delta G \propto \Delta x \propto J$, the programming current should vary in a similar manner as the variation of the synaptic plasticity ($\Delta G$ variation) with spike timing difference of connecting neurons. Such an intuitive variation of programming current variation for synaptic plasticity implementation is again a functionality offered by the decoupled ``write'' and ``read'' current paths of the proposed device structure. The programming current flows through the constant HM resistance and is not impacted by the present synaptic MTJ conductance magnitude. This results in simple peripheral circuit design as well for implementing the desired plasticity rule. In contrast, conductance change in traditional two terminal memristors depend on the history of the programming pulses.

The operating mode of the synapse, i.e. the spike transmission (``read") or the programming (``write") mode is determined by the control signal POST. The access transistors causes the isolation of the appropriate device terminals during ``write"/``read" operations. When the POST signal is deactivated, terminals $T_1$ and $T_3$ of the device are activated and spike voltage signals can be transmitted from the pre-neuron ($V_{SPIKE}$) signal through the MTJ conductance to provide an equivalent amount of synaptic current to the post-neuron circuit (connected to terminal $T_3$). When the POST signal is activated the ``write" current path through terminals $T_2-T_3$ gets activated and the device state is updated depending on the amount of synaptic current being supplied by the interfaced $M_{STDP}$ transistor. Note that the terminal $T_3$ is connected to GND during ``write" mode of operation of the device and is disconnected from the post-neuron.
\begin{figure*}
\centering
\includegraphics[width = 5in ]{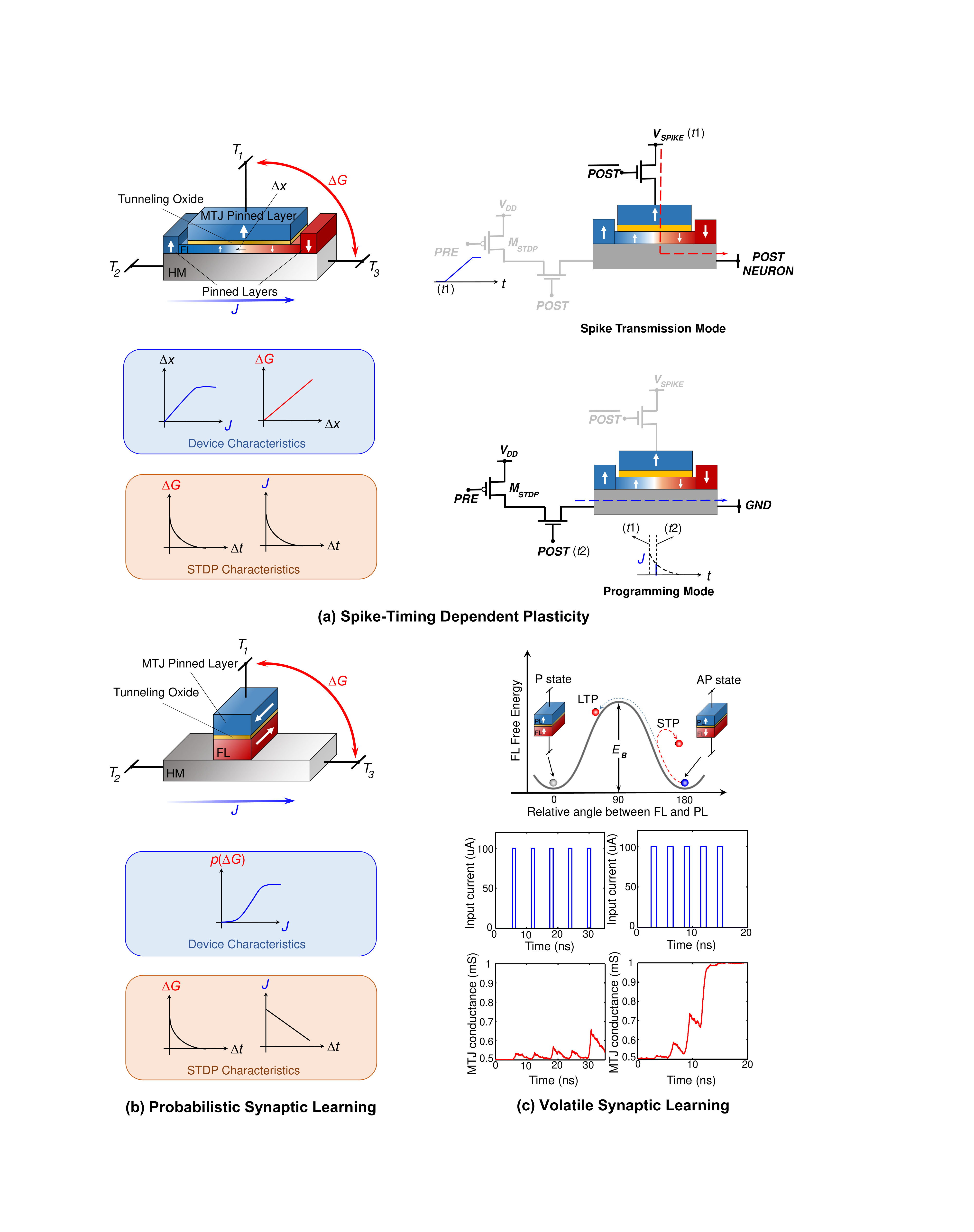}
\caption{Spike-Timing Dependent Plasticity: Magnitude of current flowing through the underlying HM, $J$, causes a proportionate displacement, $\Delta x$, in the DW position, which causes a change, $\Delta G$, in the device conductance between terminals $T_1$ and $T_3$. The device characteristics illustrate that the programming current magnitude is directly proportional to the amount of conductance change, provided the DW velocity is below the saturation value. STDP characteristics is implemented by biasing the transistor $M_{STDP}$ in subthreshold saturation regime in order to achieve the exponential current dynamics through the HM layer . The spike transmission and programming current modes are depicted in the right hand panel where the PRE and $V_{SPIKE}$ signals are activated at pre-neuron firing event at time $t_1$. POST signal, activated at post-neuron firing event at time $t_2$, samples the appropriate amount of programming current corresponding to the spike timing difference. }
\label{synapsea}
\end{figure*}

Let us now consider the learning mechanism in the spintronic device in more details. The most common learning rule dictates an exponential reduction in conductance change with increase in the value of spike timing difference. The exponential variation of current through the HM can be obtained by biasing the interfaced transistor $M_{STDP}$ in the sub-threshold regime ($V_{gs}<V_{t}$ and $V_{ds}>4U_{T}$, $V_{t}$: threshold voltage and $U_{T}$: thermal voltage) since the current flowing through the transistor will vary exponentially with the gate to source voltage. Thus, for a linear increase of the gate voltage (PRE signal) every time a pre-neuron spikes, the peripheral programming transistor will be driven from cut-off to the sub-threshold saturation region when the POST signal is activated and an appropriate programming current (magnitude varying exponentially with timing difference of pre- and post-neuron spikes) should flow through the HM. The duration of the programming current is determined by the duration of the POST signal and the magnitude is determined by the current supplied by the bias-point (PRE signal) of the $M_{STDP}$ transistor. It is worth noting here that the relationship $\Delta G \propto \Delta x \propto J$ is valid when the magnitude of the programming current $J$ remains constant during the programming duration. This is achieved by ensuring that the rise time of the gate voltage PRE of the $M_{STDP}$ transistor, or equivalently the STDP time constants, are much longer than the programming time durations (duration of POST signal) such that the current flowing through the HM of the spintronic synapse remains approximately constant. Ref. \cite{sengupta2015hybrid} considered STDP timing constants in the range of $\sim \mu s$ whereas the duration of the POST signal was $1ns$. For a linearly rising gate voltage from $0.2$ to $0.6V$ of the $M_{STDP}$ transistor (drain voltage being at $0.6V$), exponential current dynamics was observed due to transistor operation in the sub-threshold saturation regime. The linearly rising gate voltage can be easily implemented by charging a capacitor with a constant input current source everytime a pre-neuron spikes \cite{sengupta2015hybrid}. 

The discussion so far has been limited only to the implementation of the positive timing window of the STDP curve. In order to implement both the timing windows, an additional NMOS transistor is utilized in parallel to the PMOS transistor $M_{STDP}$. Two separate learning circuitries are utilized for each of the timing windows which consists of a capacitor being charged by a current source. Every-time the pre-neuron spikes, the circuit for the negative timing window is reset first such that the gate voltage of the NMOS transistor starts increasing with time. Since the drain of the NMOS transistor is negative (in order to pass current through the HM in the opposite direction for the negative timing window), the current supplied by the NMOS transistor increases as the delay of activation of the POST signal increases. In order to account for both the timing windows, the POST signal is activated after a delay of the negative timing window in order to sample the programming current contributions from the learning circuits for the positive and negative timing windows. Hence if the post-neuron spikes before the pre-neuron (negative window), the programming path will be activated during the time duration the gate voltage of the NMOS transistor is rising to pass a negative current through the device and thereby reduce the device conductance. After the duration of the negative timing window, the learning circuit for the positive timing window is reset and the POST signal is activated during this window only for a potentiation event, i.e. post-neuron ``spiking" after pre-neuron. Note that the learning circuitry which consists of the capacitor and the current source transistors can be shared across all the synapses being driven by the same pre-neuron. Discussions of crossbar arrays of such spintronic synapses for SNN implementations with on-chip learning capabilities will be discussed in the next section. Detailed operations explaining the implementation of synaptic plasticity is explained in Fig. \ref{synapsea}.

As discussed previously, the ``read" operation of the spintronic device or the synaptic scaling operation is a direct consequence of Kirchoff's law. For a constant magnitude of the spike signal, $V_{SPIKE}$, the current flowing through the synapse gets multiplied by the synaptic conductance. However, it is worth noting here that the conductance of the device is a function of the applied voltage as well. The resistance in the AP state is a much stronger function of the applied voltage than the P state and reduces by a significant amount as the applied voltage increases. Hence, higher the magnitude of the spike signal lower is the ratio of the maximum to the minimum synaptic conductance achievable. Note that higher synaptic weight ratios are desirable for achieving higher accuracy in pattern recognition workloads. Hence in order to maximize the discrimination between the two synaptic states, it is important to operate the synapses at low operating voltages less than $100mV$. This can be easily achieved by interfacing such synapses with magneto-metallic spin neurons (which inherently require low currents for switching) \cite{sengupta2015magnetic} or CMOS neurons operating in the subthreshold saturation regime \cite{indiveri2003low}. Operating the synapses at lower voltages is more important for ``non-spiking" networks since the neuron inputs need to be analog in nature. Hence the voltages applied across the synapse would be different for different inputs, thereby causing the synaptic weight to be a function of the applied input. Thus it is imperative to operate the synapses at low voltages from a functional perspective. Lower operating voltage assists in reducing the maximum ``read" current flowing through the device which, in turn, determines the device width. Assuming that the main spin torque exerted on the FL due to the ``read" current being from SOT generated by the HM, the device width can be scaled up to ensure that no DW depinning occurs for the maximum allowable magnitude of the ``read" current. The length of the synapse would be determined by the maximum number of states required from algorithm perspective. Predictive analysis performed by Sengupta $et$ $al.$ for such SOT induced plastic CoFe-Pt synapses demonstrate programming energies per synaptic event which is an order of magnitude lower than programming energies reported for a 4-bit SRAM synapse at $10nm$ technology node \cite{rajendran2013specifications}. Interestingly, analysis performed by Rajendran $et$ $al.$ revealed that although analog neuromorphic systems based on typical emerging memristive technologies will provide area benefits at scaled technologies, power consumption would be twice as high in comparison to its digital counterpart \cite{rajendran2013specifications}. This is because resistive technologies like GeSbTe \cite{jackson2013nanoscale,kuzum2011nanoelectronic}/Ag-Si \cite{jo2010nanoscale} devices are usually characterized by high threshold voltages $\sim V$ and involve much higher programming energies in the range of $\sim pJ$ and programming time durations in the range of $\sim \mu s$. Low-power on-chip learning enabled by such spintronic synapses can potentially bridge this energy in-efficiency gap. Associative memory operations using multi-domain spin-orbit torque devices has been demonstrated experimentally \cite{borders2016analogue}.

\subsubsection{Probabilistic Synaptic Learning}
\begin{figure}
\centering
\includegraphics[width = 2.5in ]{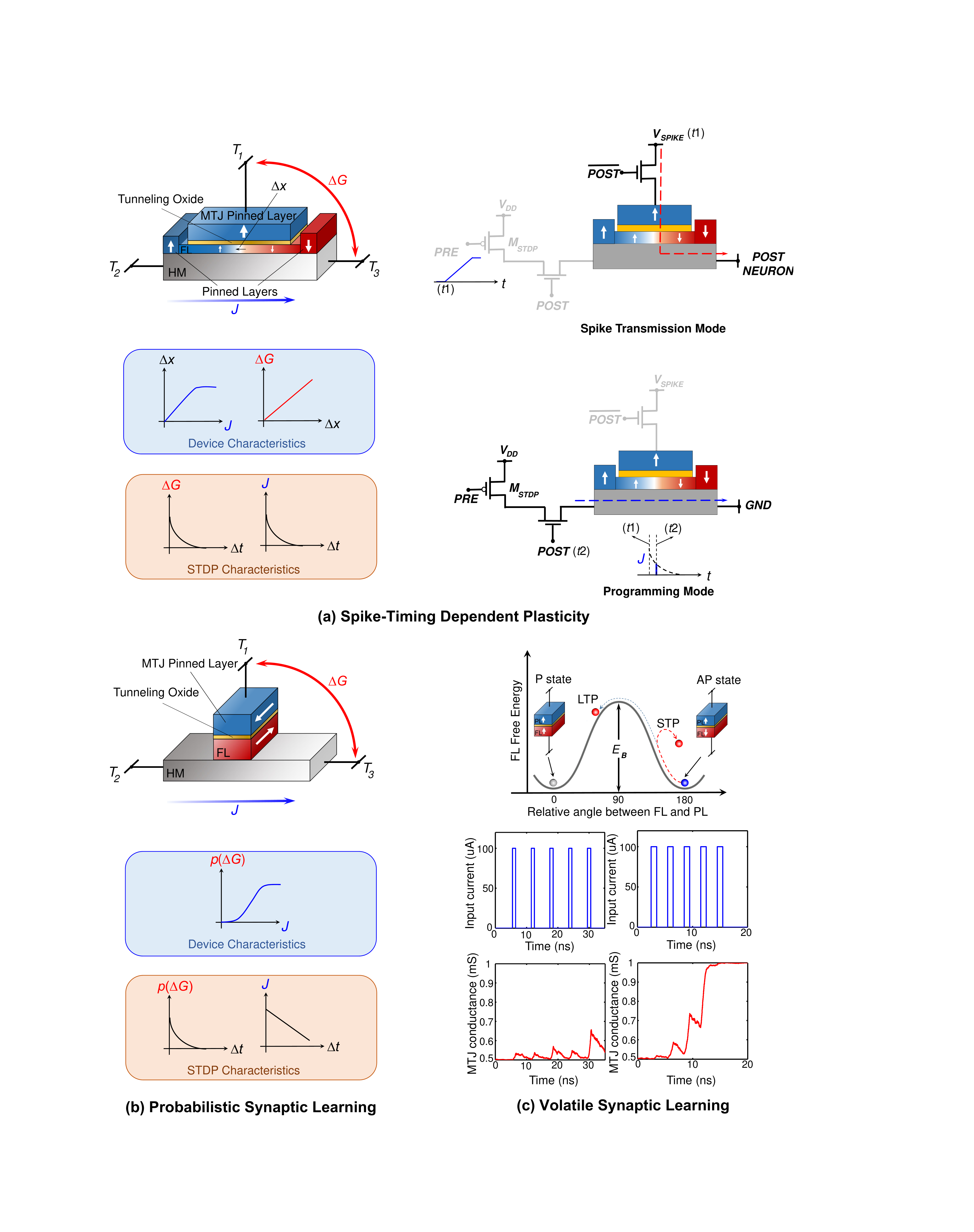}
\caption{Probabilistic STDP learning: This can be achieved in a similar fashion in mono-domain MTJ synapses by exploiting sigmoidal stochastic device switching characteristics. In the low switching probability regime (for ensuring non-greedy learning), the ``write" current reduces linearly with spike timing to emulate exponential probabilistic STDP characteristics. This is ensured by biasing $M_{STDP}$ in the saturation regime.}
\label{synapseb}
\end{figure}
The complementary version of single-bit probabilistic STDP can be similarly implemented using the single-domain MTJ-HM bilayer structures discussed previously. While Vincent $et$ $al.$ explored a simplified version of probabilistic STDP where the probability of synaptic state change was constant for positive and negative timing windows \cite{vincent2015spin}, Srinivasan $et$ $al.$ proposed crossbar architectures of such MTJ-enabled stochastic learning where the update probability varied exponentially with spike timing in accordance to original STDP formulations \cite{srinivasan2016magnetic}. As explained in Fig. \ref{synapseb}, this can be achieved by a similar framework described for the DW motion based devices where an additional interfaced transistor $M_{STDP}$, biased in the saturation regime, is driven by a linearly increasing gate voltage every time the pre-neuron spikes \cite{srinivasan2016magnetic}. Another potential advantage of probabilistic learning is below-threshold operation of devices. Since the update probability is maintained typically below 0.1 to maintain ``non-greedy" learning \cite{scholz2001micromagnetic,srinivasan2016magnetic}, operating current and voltage requirements of such devices are significantly reduced.

\subsection{Volatile Synaptic Learning}
\begin{figure}
\centering
\includegraphics[width = 3in ]{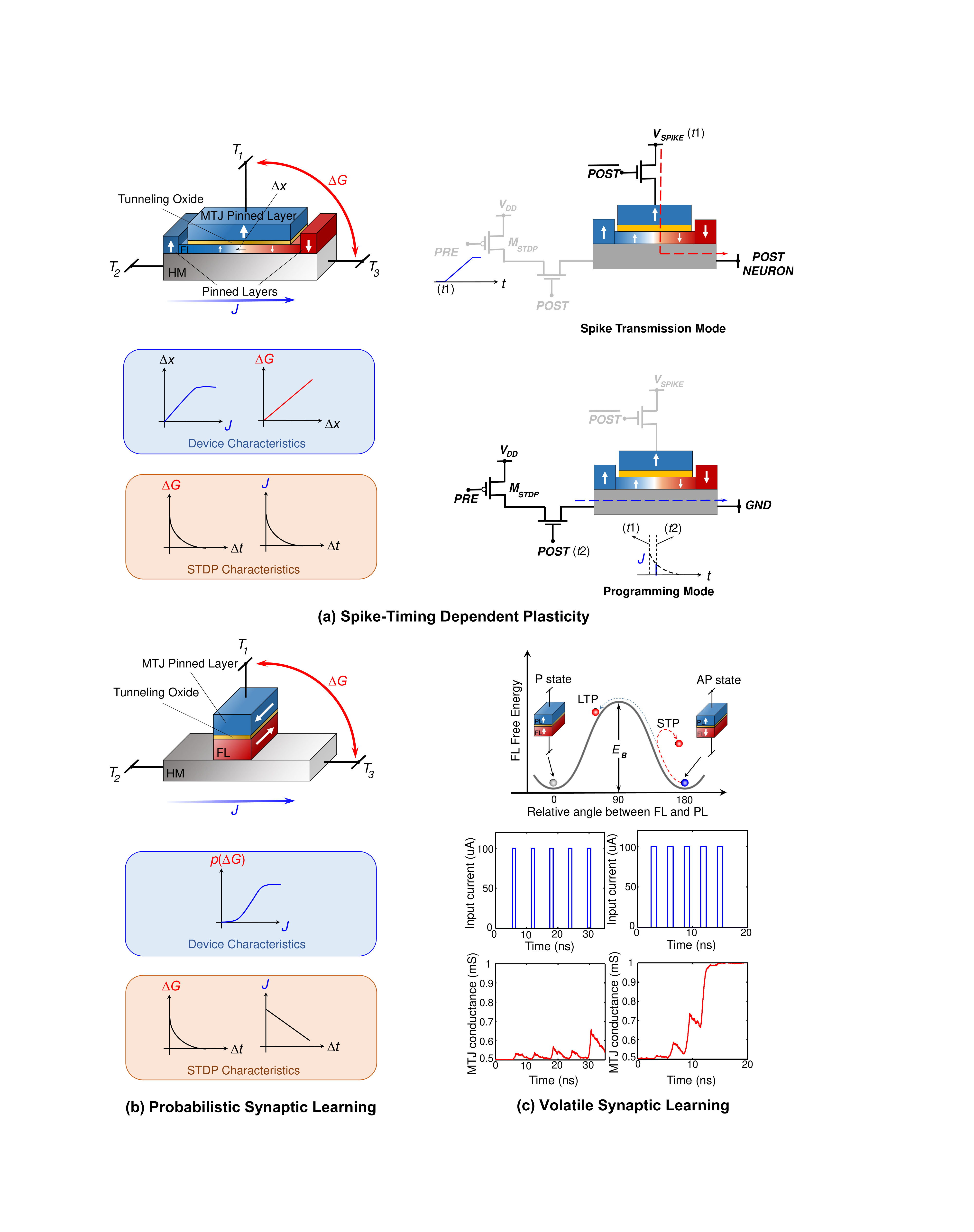}
\caption{Frequency dependent volatile synaptic learning: A mono-domain MTJ is characterized by two stable states separated by an energy barrier $E_B$. If the frequency of the input stimuli is not enough, the MTJ is unable to cross the metastable position at $90^{o}$ relative angle between FL and PL and stabilizes back to the initial magnetization state, exhibiting STP. As the stimuli frequency increases, the MTJ exhibits a much higher probability of switching to the other stable state, thereby exhibiting LTP \cite{sengupta2016short}.}
\label{synapsec}
\end{figure}
In order to implement frequency dependent volatile synaptic learning, a nanoelectronic device is required that exhibits only two stable resistive states and undergoes meta-stable state transitions whenever an input stimulus is received. The apparent spintronic device that can be directly mapped to such a functionality is the mono-domain MTJ where the spin-polarization of incoming electrons can be thought to be analogous to the release of neurotransmitters in a biological synapse. Fig. \ref{synapsec} depicts two different instances where an MTJ is subjected to five current pulses each of duration $1ns$. While the MTJ switches to the opposite state (LTP) when the time-interval of stimulation is low (3 $ns$), it is unable to switch (STP) when the stimulation time-interval is in-frequent (6 $ns$). This phenomena is due to the leaky-integrate time-varying dynamics of the magnetic FL. In the presence of an input spike (current pulse), the magnetization starts integrating (switching) towards the opposite stable magnetization state. However, in case the pulse is removed before the entire switching event can take place, the magnetization starts leaking back towards the original magnetization state. Ref. \cite{sengupta2016short} provides an in-depth discussion on STP and LTP mechanisms exhibited in such MTJ structures and demonstrates that paired-pulse facilitation (PPF: synaptic plasticity increase when a second stimulus follows a previous similar stimulus) and post-tetanic potentiation (PTP: progressive synaptic plasticity increment when a large number of such stimuli are received successively) measurements for an MTJ \cite{sengupta2016short} closely resemble those performed in frog neuromuscular junctions \cite{magleby1973effect}.

\section{Spin Based Neuromorphic Circuits and Systems}
\subsection{All-Spin Neural Networks}
\begin{figure*}
\centering
\includegraphics[width = 7.1in ]{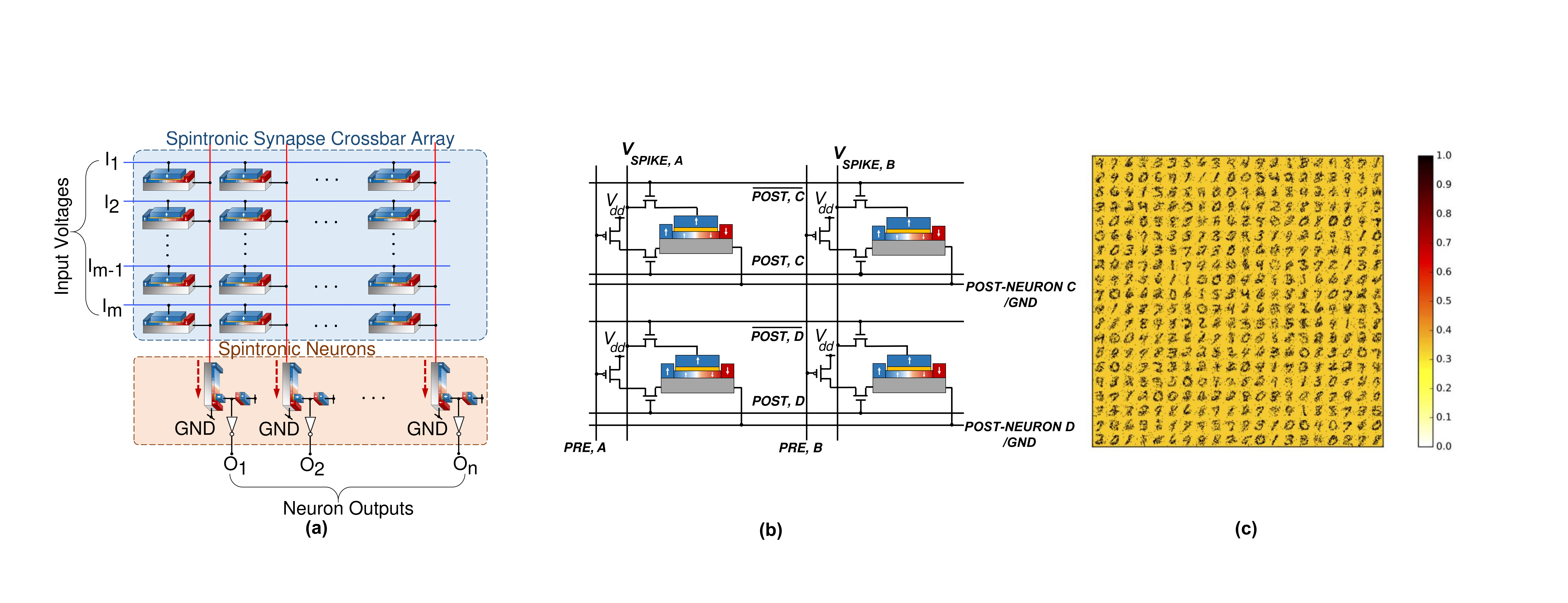}
\caption{All-Spin Neural Networks: A particular layer of a neural network with $m$ inputs and $n$ outputs can be mapped to a crossbar array of dimension $m \times n$. At a particular time-step, the rows corresponding to those inputs which have spiked are asserted a HIGH voltage level while zero voltage is applied along the rows for the ``non-spiking" inputs. Since the input ``write" resistance of the magneto-metallic spin-neurons is low, the resultant current provided by each column of the crossbar array as input to the corresponding spin-neuron equals approximately the dot-product of the neuron inputs and the corresponding synaptic weights. (b) STDP learning: Detailed hybrid spintronic-CMOS crossbar array is depicted for the implementation of STDP learning. Each spintronic synapse is interfaced with programming and access transistors. The $2\times 2$ array connects pre-neurons A and B to post-neurons C and D. (c) Digit Recognition: Learnt digit patterns in the MTJ synaptic weights of probabilistic STDP-enabled networks for 400 neurons at the end of the training phase are depicted \cite{srinivasan2016magnetic}.}
\label{crossbar}
\end{figure*}
Irrespective of the network connectivity (FCN/ CNN) the main computing kernel involved in such computing schemes can be mapped to a parallel dot-product implementation followed by neural processing. Let us begin the discussion in this section by considering spintronic synapses to be the multi-bit DW motion based device structures driving similar IF ``spiking" neurons discussed in the previous section. For this subsection, we will assume offline learning of such networks where the synaptic weights are pre-determined by backpropagation \cite{diehl2015fast} and on-chip learning functionality is not involved. Enabling on-chip intelligence in SNNs will be illustrated in the next subsection. 

The main underlying principle for implementation of the parallel-dot product computing kernel is based on the very simple and intuitive application of Kirchoff's laws. Considering a dot-product operation between $m$ inputs and $n$ outputs, the computation can be represented by a crossbar array of dimension $m \times n$ (Fig. \ref{crossbar}(a)). At each cross-point of the array, a spintronic synaptic device is present whose conductance encodes the value of the corresponding synaptic weight. Whenever a ``spike" is received at a particular input, a high voltage signal is applied along the row while a no ``spike" is represented by a low voltage signal. Assuming all the vertical lines of the array to be at ground potential, the current flowing through each crosspoint will be weighted by the synaptic conductance and get summed up along the column to provide a resultant input current (representing the dot product) to the neuron for further processing. Note that this is a major advantage of such ``in-memory" computing architectures since the synaptic weights can be stored locally in the non-volatile resistive states of the spintronic devices arranged in a crossbar fashion. In contrast, CMOS based neuromorphic architectures involve significant energy consumption due to memory leakage and memory access in order to fetch the synaptic weight values to the neural computing core for each input spike. 

In order to maintain the vertical columns at ground potential, prior work has mostly considered interfacing the crossbar arrays with analog CMOS neurons that can maintain the vertical columns at virtual ground \cite{prezioso2015training}. Note that the basic functionality that we are exploiting in the design of spintronic neuronal device structures is also that of a programmable resistor. However, the main reason such device structures are suitable for neural as well as synaptic operations is due to the decoupled nature of the ``write" and ``read" current paths. The input resistance of the device during the ``write" operation is mainly the low HM resistance and hence the synaptic input current from the crossbar array is not required to flow through the MTJ oxide. Further such magneto-metallic spin-neurons are characterized inherently by low switching current requirement thereby minimizing the terminal voltage drop across such devices. This is the main reason attributed to the usage of other two terminal resistive memories \cite{jackson2013nanoscale,kuzum2011nanoelectronic,jo2010nanoscale} primarily as synaptic devices. Interfacing such two terminal memristive crossbar arrays with two terminal memristive neurons would be potentially difficult resulting in erroneous dot product computation since the vertical columns of the array would be no longer maintained at ground potential (due to the high threshold voltages and resistances of such memory technologies). In addition to providing the flexibility of implementing neuronal and synaptic devices by the same technology, spintronic neurons enable low power operation of the spintronic crossbar array due to low switching current requirements of such magneto-metallic devices. In contrast, analog CMOS neuron implementations typically require the crossbar arrays to be run at a much higher voltage. 

Let us now consider the operation of the crossbar array in more details. Each time-step of SNN operation consists of a neuron ``write" cycle followed by the ``read" and ``reset" cycles. In order to implement bipolar weights, two rows ($V_{i+}$ and $V_{i-}$) are used for each input $V_i$. When the input $V_i$ assumes a logic value of \textquoteleft 0\textquoteright (no ``spike"), then \textquoteleft 0\textquoteright \ voltage level is applied to both the inputs. However, when $V_i$ assumes a logic value of \textquoteleft 1\textquoteright (``spike"), then voltage $+\Delta V$ (less than $100mV$) is applied to the row corresponding to $V_{i+}$ and $-\Delta V$ is applied to the row corresponding to $V_{i-}$. If the weight $w_{i,j}$ for the $j$-th neuron and input $V_i$ is positive, then the conductance corresponding to $V_{i+}$ is programmed to $G_{i,j+}=w_{i,j}.G_{o}$ ($G_{o}$ is the mapped conductance for unity synaptic weight), while the conductance, $G_{i,j-}$ corresponding to $V_{i-}$ is programmed to high OFF resistive state and vice versa. Let us consider the input conductance of the spintronic neuron during the ``write'' operation (mainly the HM conductance of the neuron) to be $G_s$ and the voltage drop across the neuron to be $V_s$. Equating the current supplied by the resistive synapses to the current flowing through the neuron, we get $\sum\limits_{i}(G_{i,j+}.(V_{i+}-V_s)+G_{i,j-}.(V_{i-}-V_s))=G_s.V_s$ which indicates that the net synaptic current supplied to the spintronic neuron is given by, 
\begin{equation}
\begin{aligned}
I_j&=G_s.V_s \\
&=\frac {G_s. \sum\limits_{i}(G_{i,j+}.V_{i+}+G_{i,j-}.V_{i-})} {G_s+ \sum\limits_{i}(G_{i,j+}+G_{i,j-})} \\
&=\frac{\sum\limits_{i}(G_{i,j+}.V_{i+}+G_{i,j-}.V_{i-})}{1+\gamma}
\end{aligned}
\end{equation}

As mentioned previously, it is imperative to run spintronic crossbar arrays at low operating voltages from functionality viewpoint. However, lower the operating voltage, higher is the range of synaptic conductances (which can be appropriately tuned by choosing a proper value of MTJ oxide thickness) required to ensure sufficient current requirement for DW displacement from one edge to another in the FM of the spintronic neurons. Hence lower crossbar operating voltage results in the increment of the ratio, $\gamma =\sum\limits_{i}(G_{i,j+}+G_{i,j-})/G_{s}$, which in turn, results in non-ideal operation of the neuron. In order to ensure that $\gamma << 1$ for a given crossbar operating voltage, the duration of the ``write'' cycle can be adjusted accordingly since the current required to achieve a specific DW displacement scales linearly with the duration of the ``write'' current. The output signals of the inverters from a particular array can be stored in a latch and used to communicate input signals to the fan-out neurons being implemented in the crossbar array for the succeeding stage. Note that the latched neuron outputs can be also used to drive input rows of the same crossbar array (inputs for the next time-step) to implement recurrent neuron connections in RNN architectures.

Ref. \cite{sengupta_allspinsnn} evaluated the circuit-level performance of such an All-Spin SNN based design against a baseline CMOS implementation at $45nm$ technology node for a benchmark digit recognition problem on the MNIST dataset \cite{lecun1998gradient}. The neural network connectivity used was a typical Deep CNN architecture (28x28-12c5-2s-64c5-
2s-10o). The ``write" cycle duration was optimized to $2ns$ to ensure competitive accuracy over the testing set ($\sim 98.5\%$). Simulation studies of the entire network indicate that the proposed spintronic design can potentially achieve $250\times$ improvement in energy consumption and $56\times$ improvement in EDP over a baseline digital CMOS implementation in commercial $45nm$ technology \cite{sengupta_allspinsnn}. Note that this is a circuit level comparison work where the IF ``spiking" neurons are implemented using digital adders and a comparator. A pipelined CMOS implementation was considered with power-gating to exploit the event-driven nature of ``spiking" networks. Memory access overhead for CMOS based architectures would further increase the energy benefits offered by such All-Spin SNN designs.

While the above discussion considered offline trained Deep CNNs driven by deterministic DW motion based IF ``spiking" neurons, similar SNN networks can be trained for stochastic ``spiking" neurons enabled by single-domain MTJs. Ref. \cite{sengupta2016probabilistic} explored an approach of training deep CNNs with sigmoid transfer function neurons using backpropagation and subsequently utilizing the offline trained weights to implement an SNN where the neurons generate output spikes at each time-step using sigmoid probability distribution functions. Although a sigmoid neuron function in the ``non-spiking" domain is not exactly equivalent to a probabilistic sigmoid ``spiking" neuron transfer function, authors in Ref. \cite{sengupta2016probabilistic} demonstrate that the two implementations can be reasonably close to achieve similar recognition accuracies as the offline trained ``non-spiking" network. The advantages of such an approach is driven solely by the fact that complex neural operations (like sigmoid transfer functions) required to achieve high recognition accuracies can be now implemented by simple device structures consisting of mono-domain magnets by leveraging the underlying device stochasticity. Design considerations for the implementation of such networks also rely on the proper choice of the ``write" cycle duration as in the previous case. The stochastic sigmoid characteristics of a nanomagnet undergoes more dispersion as the ``write" cycle duration reduces \cite{sengupta2016probabilistic}. Since the range of synaptic conductances in the crossbar array increases to provide more input current to the neuron as the ``write" cycle duration decreases (due to increased dispersion of the MTJ probability switching characteristics in response to the input current), increment in ``write" cycle duration is necessary to reduce the impact of the non-ideality parameter $\gamma$ discussed previously. However, as the ``write" cycle duration increases, the sigmoid switching characteristics becomes more steeper resulting in increased sensitivity to process and temperature variations. Ref. \cite{sengupta2016probabilistic} considered the implementation of such stochastic Deep SNN networks (28x28-6c5-2s-12c5-2s-10o) on the MNIST dataset \cite{lecun1998gradient}. For an optimal ``write" cycle duration of $0.5ns$ and 50 time-steps of operation of the network per input image (to achieve competitive classification accuracies), the MTJ enabled stochastic SNN was evaluated to be $20 \times$ more energy efficient than the baseline CMOS implementation.

\subsection{STDP Learning}
For clarity, the learning circuitry for SNN was omitted in the above discussion. To better understand device, circuit and system level efficiencies with spin-synapses in the context of learning, let us consider the STDP-enabled single layer SNNs discussed in Section III.4. The network functionality can be mapped to a crossbar array as shown in Fig. \ref{crossbar}(b) where spike signals transmitted along the rows from the pre-neurons get summed up along the columns to the post-neurons. The spintronic synapses are programmed only when the post-neuron spikes (with a delay of the negative timing window) and are switched off from the post-neuron circuit during the programming phase using the POST control signal. Each cross-point consists of a spin-synapse interfaced with access transistors and $M_{STDP}$ transistor. An additional programming transistor is also present at each cross-point for the negative timing window but is not shown in Fig. \ref{crossbar}(b) for illustrative purposes. Note that the spin-synapses at each cross-point can be either the DW motion or the single-domain based MTJ device depending on the nature of STDP functionality being implemented. The inhibitory functionality in such networks can be implemented by an additional row in the crossbar array that is driven by a negative voltage. The row should be activated whenever any of the neurons generate an output spike to prevent multiple neurons from learning the same pattern. The post-neurons can be subthreshold CMOS neurons \cite{indiveri2003low} or MTJ based ``spiking" neurons \cite{sengupta2015magnetic}. Unsupervised multi-bit STDP learning with MTJ ``spiking" neurons has been demonstrated in Ref. \cite{sengupta2015magnetic}. Probabilistic STDP based on spintronic synapses in such single layer networks have been also demonstrated in Ref. \cite{srinivasan2016magnetic} and have been able to achieve $\sim 80\%$ recognition accuracy over the MNIST \cite{lecun1998gradient} training set for a set of 225 excitatory neurons. Such networks have been shown to achieve competitive recognition accuracies by increasing the neuron count beyond 1000. Fig. \ref{crossbar}(c) depicts learnt digit patterns in the MTJ synaptic weights of probabilistic STDP-enabled networks for 400 neurons at the end of the training phase. Interested readers are referred to Ref. \cite{gopal2017} for an overview of All-Spin Stochastic SNNs where stochastic synaptic learning is accomplished by probabilistic neural inference, both enabled by single-domain MTJ devices. It is worth noting here that such stochastic computing paradigms are equally valid for magnets scaled to the super-paramagnetic regime. However, appropriate circuit considerations need to be accounted for due to the telegraphic switching behavior of such low barrier magnets \cite{liyanagedera2017magnetic}.

Inspired by the concept of Long-Term and Short-Term Memory, researchers have proposed alternative significance driven network architectures that attempt to improve on the learning convergence of such STDP-enabled SNNs. Instead of considering that each neuron receives a resultant synaptic input from a single crossbar, the computation can be split up into two separate crossbars, namely the Long-Term (LT) and Short-Term (ST) arrays. The LT array is characterized by STDP timing constants that are comparatively smaller than the ST array. This can be easily achieved by tuning the capacitance value in the peripheral learning circuitry mentioned earlier that generates the linearly increasing gate voltage of the $M_{STDP}$ transistor. Consequently, the LT array learns input representations that are very strongly correlated in time in comparison to the ST array. The ST array acquires the moderately-correlated general features in the input pattern. Hence, while inference (``read" operation) the LT array is attributed a higher significance by driving the rows at a higher voltage in comparison to the ST array. Ref. \cite{srinivasan2016magnetic} demonstrated the efficiency of such significance driven architectures by performing an iso-accuracy analysis for the case of a single crossbar (with 400 neurons and 313600 synaptic units) versus the LT-ST crossbar arrays (225 neurons and 352800 synaptic units). Faster training convergence in the LT-ST case resulted in $\sim 2\times$ reduction in the total expended programming energy during the training process while incurring minimal area penalty.

\subsection{System Level Benchmarking}
\begin{figure}
\centering
\includegraphics[width =3in ]{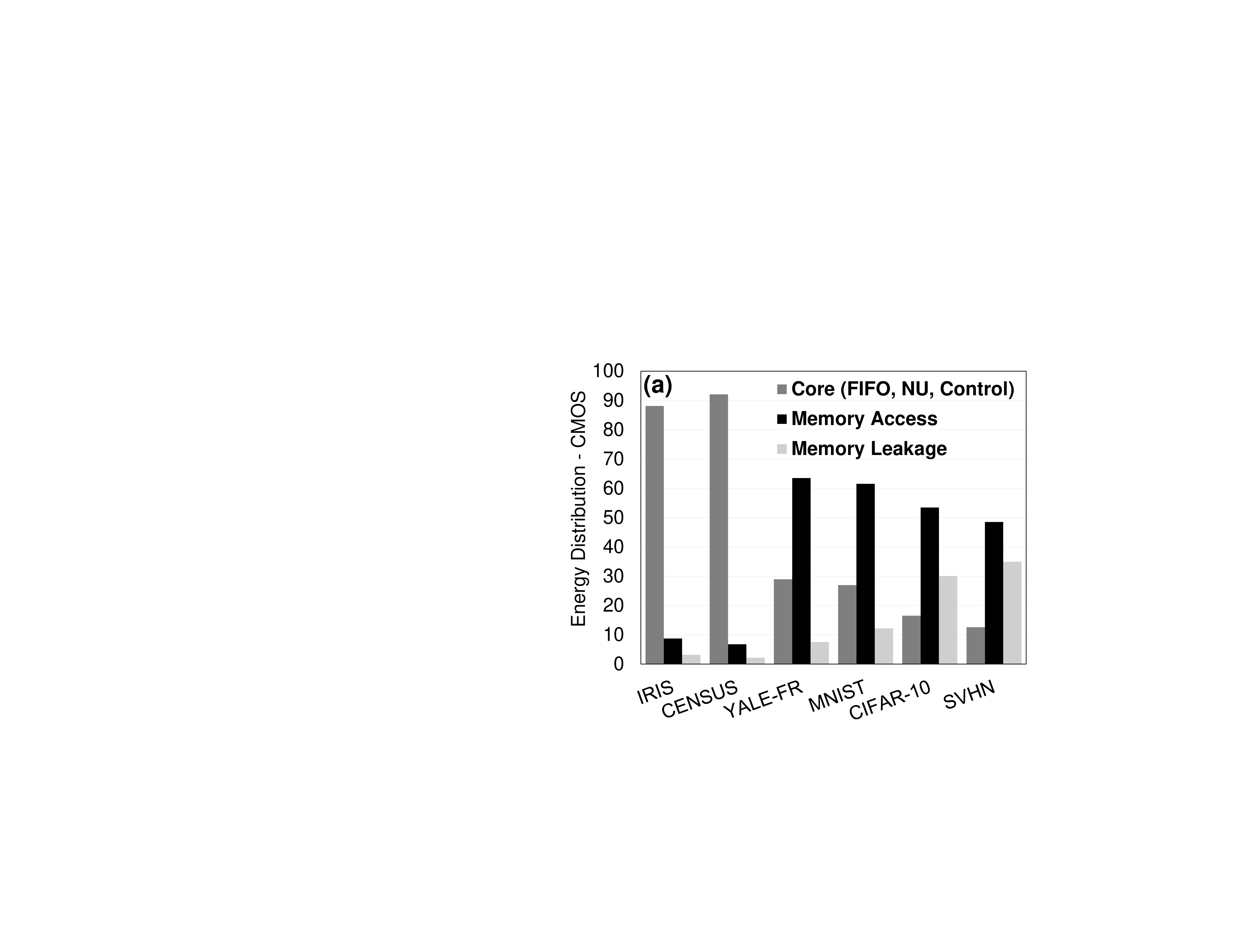}
\includegraphics[width =3in ]{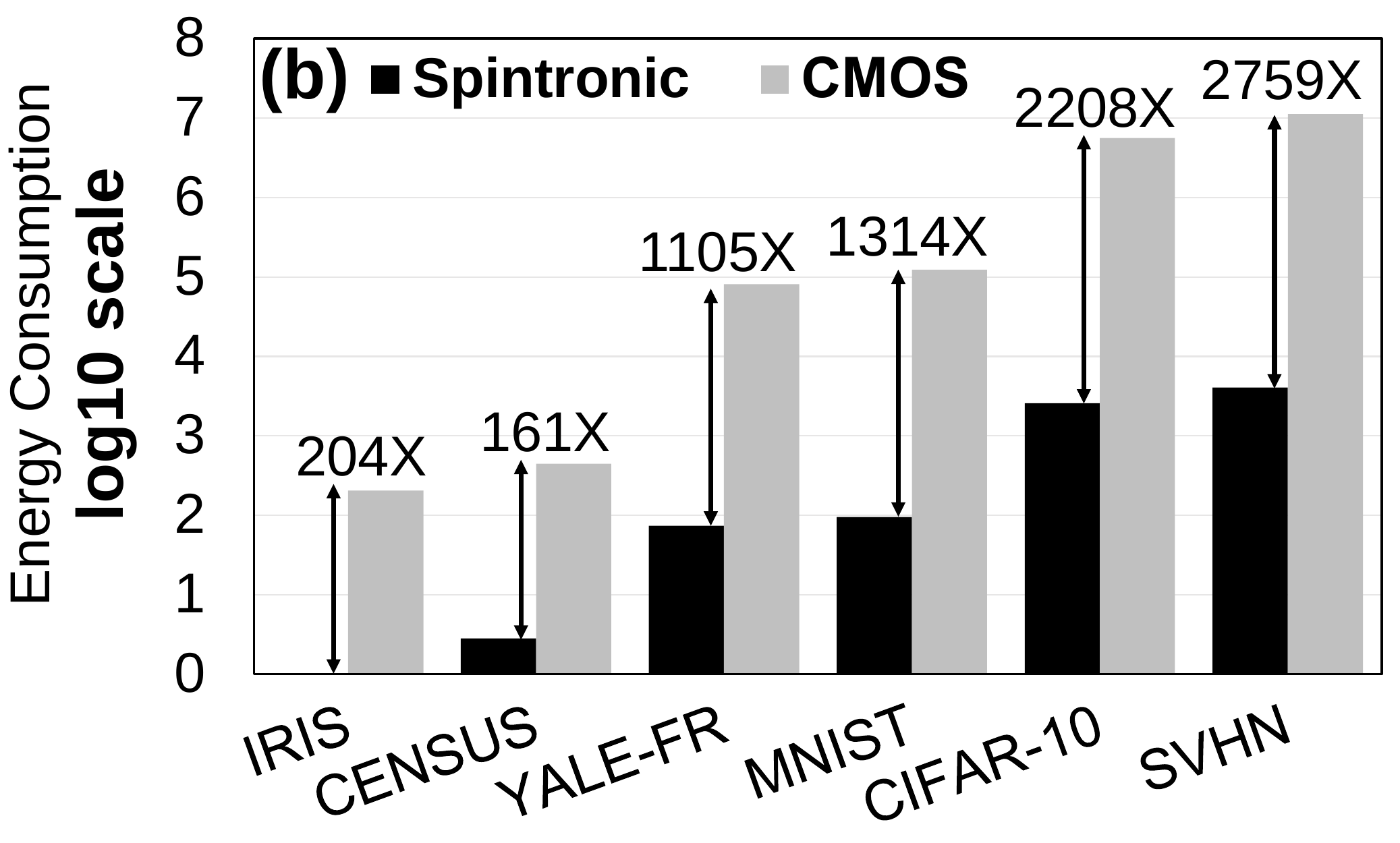}
\includegraphics[width =3in ]{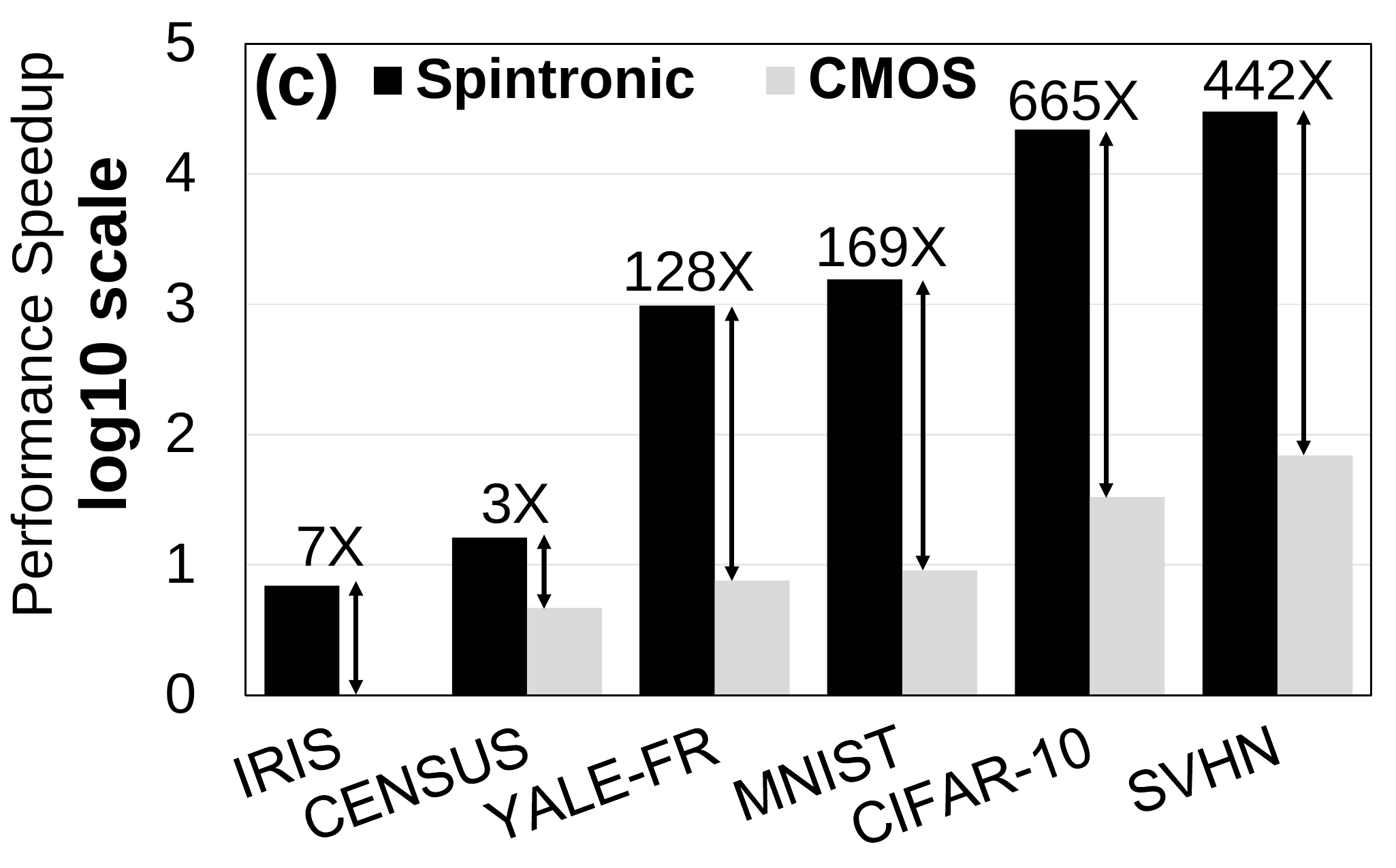}
\caption{(a) Energy distribution profile for the CMOS architecture. (b) Energy consumption comparison between Spintronic and CMOS architectures. (c) Performance speedup comparison between Spintronic and CMOS architectures \cite{benchmarking}. The benchmark suite consists of the following applications: (i) Flower Species Recognition (IRIS dataset \cite{uci}), (ii) Census data analysis (ADULT dataset \cite{uci}), (iii) Face recognition (YALE dataset \cite{yale}), (iv) Digit recognition (MNIST dataset \cite{lecun1998gradient}), (v) Object classification (CIFAR-10 dataset \cite{krizhevsky2009learning}) and (vi) House Number Recognition (SVHN dataset \cite{netzer2011reading}).}
\label{results}
\end{figure}
Authors in Ref. \cite{benchmarking} performed a rigorous system-level benchmarking of a reconfigurable neuromorphic architecture based on such All-Spin SNNs. A benchmark suite of 6 recognition problems ranging in network complexity from 10k-7.4M synapses and 195-9.2k neurons was used \cite{benchmarking}. The reconfigurable spintronic architecture was based on $32 \times 32$ sized crossbar arrays and was time-multiplexed using global control logic for proper functionality. The baseline CMOS architecture consists of an SRAM to store the trained weights and a computation core to fetch the weights from SRAM and perform the inner-product and neuron computations. 

A hybrid device-circuit-architecture co-simulation framework used in Ref. \cite{benchmarking} reveal that the All-Spin SNN architecture can potentially achieve $204 - 2759 \times$ improvement in energy consumption while achieving $3 - 665 \times$ performance speedup in comparison to the CMOS baseline implementation (Fig. \ref{results}). Significant proportion of energy is expended in memory access and memory leakage in comparison to the core computation and this proportion increases with increased problem complexity. Additionally, the access latency increases with increasing memory size, thereby causing a proportionate increase in the memory leakage energy. On the other hand, ``in-memory" spintronic crossbar arrays offer better crossbar utilization with increment in network size. It is worth noting here that the network-level analysis and results are based on a predictive simulation framework that consisted of device-level modelling calibrated to experimental measurements.

We would like to conclude this section by mentioning that there has been preliminary investigations to address concerns of device and circuit variations and non-idealities in such resistive crossbar arrays \cite{querlioz2013immunity}. The major advantage is that such neuromorphic algorithms are inherently resilient to reasonable approximations/errors in the neuron and synaptic units. Further, unsupervised learning is expected to enable adaptive learning in networks by taking into account the inherent non-idealities or variations in the devices and circuits.

\section{Outlook}
Spin-based neuromorphic computing is currently a technologically evolving field. While preliminary experiments are being performed that provide proof-of-concepts for the various proposals mentioned in this article, a long and interesting path lies ahead for the realization of such All-Spin neuromorphic computing platforms. Experimental demonstration of full network-level synaptic learning and neural inference based on spintronic devices remains to be explored. Innovations are still required not only at the device level (for instance, achieving deterministic DW motion or fabricating scaled nanomagnets) but also at the algorithm level to exploit the underlying device physics of spin-devices. Nevertheless, such devices offer immense possibilities towards the realization of energy-efficient cognitive processors. As device dimensions start scaling, probabilistic neuromorphic computing platforms (that are inherently more ``brain-like") leveraging the resultant device stochasticity will also start playing an important role. In conclusion, this article serves to provide a holistic review of various neural and synaptic functionalities that can be potentially implemented in spintronic devices. We believe that this article will stimulate efforts for the realization of All-Spin neuromorphic computing paradigms enabled with on-chip unsupervised cognitive learning capabilities.

%

\section*{Acknowledgements}

The work was supported in part by, Center for Spintronic Materials, Interfaces, and Novel Architectures (C-SPIN), a MARCO and DARPA sponsored StarNet center, by the Semiconductor Research Corporation, the National Science Foundation, Intel Corporation and by the US Department of Defense Vannevar Bush Faculty Fellowship.

\end{document}